\documentclass[twocolumn]{aastex631}
\usepackage{soul}
\usepackage{comment}
\usepackage{multirow}
\shorttitle{Hard X-rays in Kepler's SNR}
\shortauthors{Sapienza et al.}
\graphicspath{{./}{figures/}}

\newcommand{\nus}{\textit{NuSTAR }}
\newcommand{\xmm}{\textit{XMM-Newton}}

\begin{document}

\title{A spatially resolved study of hard X-ray emission in Kepler's SNR:\\ indications of different regimes of particle acceleration}

\author{Vincenzo Sapienza}
\affiliation{Dipartimento di Fisica e Chimica E. Segr\`e, Universit\`a degli Studi di Palermo, Piazza del Parlamento 1, 90134, Palermo, Italy}
\affiliation{INAF-Osservatorio Astronomico di Palermo, Piazza del Parlamento 1, 90134, Palermo, Italy}

\author{Marco Miceli}
\affiliation{Dipartimento di Fisica e Chimica E. Segr\`e, Universit\`a degli Studi di Palermo, Piazza del Parlamento 1, 90134, Palermo, Italy}
\affiliation{INAF-Osservatorio Astronomico di Palermo, Piazza del Parlamento 1, 90134, Palermo, Italy}

\author{Aya Bamba}
\affiliation{Department of Physics, The University of Tokyo, 7-3-1 Hongo, Bunkyo, Tokyo 113-0033, Japan}
\affiliation{Research Center for the Early Universe, School of Science, The University of Tokyo, 7-3-1 Hongo, Bunkyo-ku, Tokyo 113-0033, Japan}

\author{Satoru Katsuda}
\affiliation{Graduate School of Science and Engineering, Saitama University, 255 Shimo-Okubo, Sakura, Saitama 338-8570, Japan}

\author{Tsutomu Nagayoshi}
\affiliation{Graduate School of Science and Engineering, Saitama University, 255 Shimo-Okubo, Sakura, Saitama 338-8570, Japan}

\author{Yukikatsu Terada}
\affiliation{Graduate School of Science and Engineering, Saitama University, 255 Shimo-Okubo, Sakura, Saitama 338-8570, Japan}

\author{Fabrizio Bocchino}
\affiliation{INAF-Osservatorio Astronomico di Palermo, Piazza del Parlamento 1, 90134, Palermo, Italy}

\author{Salvatore Orlando}
\affiliation{INAF-Osservatorio Astronomico di Palermo, Piazza del Parlamento 1, 90134, Palermo, Italy}

\author{Giovanni Peres}
\affiliation{Dipartimento di Fisica e Chimica E. Segr\`e, Universit\`a degli Studi di Palermo, Piazza del Parlamento 1, 90134, Palermo, Italy}
\affiliation{INAF-Osservatorio Astronomico di Palermo, Piazza del Parlamento 1, 90134, Palermo, Italy}



\begin{abstract}

Synchrotron X-ray emission in young supernova remnants (SNRs) is a powerful diagnostic tool to study the population of high energy electrons accelerated at the shock front and the acceleration process.
We performed a spatially resolved spectral analysis of \nus\ and \xmm\ observations of the young Kepler's SNR, aiming to study in detail its non-thermal emission in hard X-rays.
We selected a set of regions all around the rim of the shell and extracted the corresponding spectra. 
The spectra were analyzed by adopting a model of synchrotron radiation in the loss-limited regime, to constrain the dependence of the cutoff energy of the synchrotron radiation on the shock velocity.
We identify two different regimes of particle acceleration, characterized by different Bohm factors.
In the north, where the shock interacts with a dense circumstellar medium (CSM), we found a more efficient acceleration than in the south, where the shock velocity is higher and there are no signs of shock interaction with dense CSM.
Our results suggest an enhanced efficiency of the acceleration process in regions where the shock-CSM interaction generates an amplified and turbulent magnetic field. 
By combining hard X-ray spectra with radio and $\gamma-$ray observations of Kepler's SNR, we modelled the spectral energy distribution. In the light of our results we propose that the observed $\gamma-$ray emission is mainly hadronic, and originates in the northern part of the shell.
\end{abstract}



\section{Introduction} \label{sec:intro}

Blast wave shocks in supernova remnants (SNRs) are sites of particle acceleration and are believed to be the primary source of galactic cosmic rays (CRs).
For shocks in SNRs, the main acceleration mechanism is the diffusive shock acceleration (DSA, \citealt{1978MNRAS.182..147B}, \citealt{1977ICRC...11..132A} and \citealt{1978ApJ...221L..29B}).
The first evidence for high-energy ($E>10^{12}$ eV) electrons accelerated in SNR shocks came with the detection of non-thermal X-ray emission of SN 1006 \citep{1995Natur.378..255K}.
As a matter of fact the study of X-ray synchrotron emission of SNRs can provide helpful insights about the acceleration process, such as the shape of the electron energy distribution and the mechanisms that limit the maximum energy that electrons can reach.
Different mechanisms can be invoked to limit the maximum electron energy in the acceleration process (\citealt{2008ARA&A..46...89R}); for example, it can be limited by radiative losses (\textit{loss-limited} scenario) or by the finite  acceleration time available (\textit{age-limited} scenario).

Kepler's SNR owes its name to Johannes Kepler, who extensively studied its parent supernova (SN~1604).
This remnant has a roughly spherical shape with an angular radius of approximately $1.8'$ with two characteristic protrusions (also called ``ears"), one located in the south-east of the shell and the other located in the north-west.
The SNR is very likely the result of a type Ia SN \citep{1999PASJ...51..239K}. 
\citet{2007ApJ...668L.135R} found that Kepler's SNR  is interacting with nitrogen-rich circumstellar medium (CSM) in the north and suggest a single-degenerate scenario for the explosion (with the companion possibly being a runaway AGB star, see \citealt{1987ApJ...319..885B}, \citealt{2006ApJ...649..779V}, \citealt{2012A&A...537A.139C} and \citealt{2021ApJ...915...42K}),  albeit there is no evidence for a survived companion star (\citealt{2014MNRAS.440..387K}, \citealt{2018ApJ...862..124R}).

\citet{1999AJ....118..926R} derived a distance of $4.8\pm 1.4$ kpc, based on H I absorption from radio observations, while \citet{2008A&A...488..219A} suggested a lower limit of 6.4 kpc motivated by the lack of a detectable $\gamma$-ray flux. 
However, recent estimates based on proper motion measurements derived a distance $d = 5.1_{-0.7}^{+0.8}$ kpc \citep{2016ApJ...817...36S}.
We then adopt $d=5$ kpc throughout this paper.

Prominent particle acceleration in Kepler's SNR is testified by its energetic non-thermal emission. 
The detection of GeV $\gamma-$ray emission from Kepler's SNR was recently presented by \citet{2021ApJ...908...22X} and interpreted as a signature of hadronic emission. 
Similar conclusions were reported by \citet{2022arXiv220105567A} who propose the hadronic emission to originate in the northern part of the shell, while synchrotron and Inverse Compton emission are interpreted as originating in the southern regions.
\citet{2021arXiv210711582P} reported the detection of Very High Energy (VHE) $\gamma$-ray emission from Kepler's SNR with the H.E.S.S. telescope.

The presence of non-thermal X-ray emission in Kepler's SNR was first discovered in its south-eastern region by \citet{2004A&A...414..545C}, using an \xmm\ observation. 
\citet{2007ApJ...668L.135R} conducted a spectral analysis in several regions of Kepler's SNR confirming that some of them are dominated by synchrotron radiation.
Recently, \citet{2021PASJ...73..302N} reported the first robust detection of hard X-ray emission, in the 15-30 keV band, by analyzing a \emph{Suzaku} HXD observation. 
Several spatially resolved studies found that the roll-off frequency of the synchrotron radiation in Kepler's SNR lies in the range $\nu_r \sim 1 - 8 \times 10^{17}$ Hz (\citealt{2004A&A...414..545C}, \citealt{2005ApJ...621..793B}, \citealt{2021PASJ...73..302N}).
One can estimate if the cutoff energy ($E_{max}$) of the synchrotron emitting electrons is loss-limited or time-limited by comparing the time-scale for synchrotron losses ($\tau_{sync}$) with the age of the remnant ($t_{age}$ = 418 yrs).
The timescale for synchrotron cooling is, 
\begin{equation}
    \tau_{sync}\approx 1700 \bigg(\frac{h\nu_{r}}{1 \text{ eV}}\bigg)^{-\frac{1}{2}} \bigg(\frac{B}{100\text{ } \mu\text{G}}\bigg)^{-\frac{3}{2}} \text{ yr}.
    \label{eq:tsync}
\end{equation}
On the basis of the one-zone model of broad-band Spectral Energy Distribution (SED), \citet{2021PASJ...73..302N} adopted a magnetic field of $\sim 40$ $\mu\text{G}$ and a roll-off frequency  $\nu_r = 1\times 10^{17}$ Hz, corresponding to $\tau_{sync}\sim 330$ yrs.
Non-linear DSA predicts an amplification of the magnetic field strength as a result of the flux of kinetic energy of the cosmic rays streaming ahead of the shock \citep{2004MNRAS.353..550B}. Estimates of the magnetic field strength in Kepler's SNR, based on the thickness of the X-ray synchrotron filaments, provide values in the range 170-250 $\mu$G (\citealt{2005A&A...433..229V}, \citealt{2006A&A...453..387P}, \citealt{2012A&A...545A..47R}, \citealt{2021ApJ...917...55R}).
Assuming a value of magnetic field of 170 $\mu$G and the roll of frequency measured by \citet{2021PASJ...73..302N} we obtain $\tau_{sync}\sim 30$ yrs.
In any case, the synchrotron cooling time is always lower than the age of Kepler's SNR and we can therefore consider the loss-limited scenario as the most appropriate for this source.

\citet{2021ApJ...907..117T} measured the cutoff photon energy in different regions of several SNRs, Kepler's SNR among them, by describing the non-thermal X-ray emission with the loss-limited model proposed by \citet{2007A&A...465..695Z}.
In this model, the cutoff photon energy ($\epsilon_0$) is related to the shock speed, $v_{sh}$, through $\epsilon_0 \propto v_{sh}^2 \eta^{-1}$ where $\eta$, or Bohm diffusing factor, is the ratio between the diffusion coefficient and $c\lambda/3$ (where $\lambda$ is the Larmor radius, the minimum value $\eta=1$ corresponding to the Bohm limit) and is strongly related to the turbulence of the magnetic field, which scatter the charged particles.
\citet{2021ApJ...907..117T} studied the dependence of $\epsilon_0$ on $v_{sh}$, with a spatially resolved spectral analysis in order to estimate $\eta$ in different remnants.
However, the spatially resolved analysis of Kepler's SNR lacks of the hard part of the spectrum and the $\epsilon_0 - v_{sh}$ plot shows a clear trend only for synchrotron dominated regions, while no correlation can be found for other regions.
\citet{2015ApJ...814..132L} performed a similar analysis using a deep \nus\ observation of Tycho's SNR.
They found that in Tycho's SNR the highest energy electrons are accelerated at the fastest shocks, with a steep dependence of the roll-off frequency on the shock speed.

In this paper, we present the first analysis of archive \nus\ observations of Kepler's SNR. 
We exploit the high sensitivity of the \nus\ telescope to study the morphological and spectral properties of the hard X-ray emission.
We also perform a spatially resolved measurement of the cutoff energy of the synchrotron radiation, combining the \nus\ data with an \xmm\ observation, which allows us to get physical insights on the origin of non-thermal emission. 
We describe the data reduction in Sect. \ref{sec:methods}.
Section \ref{sec:results} is dedicated to  the results of image and spectral analysis. 
Discussion and conclusions are presented in Sect. \ref{sec:conclusions} and Sect. \ref{conc}, respectively.

\section{Data reduction}\label{sec:methods}

\subsection{\nus\ observation}

The \nus\ observation of Kepler's SNR was performed from October 7, 2014 for an exposure time of 246 ks (Obs. ID: 40001020002, PI: F. Harrison), with pointing coordinates $\alpha_{J2000} = 17^h30^m36.4^s$ and $\delta_{J2000} = -21$°$30' 13''$.
We processed the data using \texttt{nupipeline} of the \textit{NuSTAR Data Analysis Software} (NuSTARDAS version 2.0.0 with CALDB version 20210202) included in HEAsoft version 6.28.

The maps shown in this paper were obtained by summing the photon counts detected in each pixel by the two Cadmium-Zinc-Telluride (CZT) detectors Focal Plane Module A and B (FPMA and FPMB) in a given energy band.
We performed a spatially resolved spectral analysis for both FPMA and FPMB by extracting the spectra from different regions of the remnant using the \texttt{nuproducts} pipeline for an extended source. Spectra were rebinned to have at least 25 counts per bin. For each region, FPMA and FPMB spectra were fitted simultaneously.
We used the \texttt{nuproducts} pipeline to produce the redistribution matrix file (RMF) and the ancillary response file (ARF), and to extract the background spectrum.
For the background, we selected an extraction region for each spectrum, outside of the shell and in the same chip as the source extraction region.

\subsection{\xmm\ Observation}

We complemented \nus\ data analysis with the analysis of an \xmm\ European Photon Imaging Camera (EPIC) observation of Kepler's SNR, performed from March 19, 2020 for an exposure time of 140 ks (Obs. ID: 0842550101, PI: T. Sato). The observation has pointing coordinates $\alpha_{J2000} = 17^h30^m36.9^s$, $\delta_{J2000} = -21$°$30' 01.1''$, and was performed with the thick filter, in full frame mode for the EPIC MOS cameras, and in large window mode for the EPIC pn camera.

We processed the Observation Data Files (ODF) using the \texttt{emproc} and the \texttt{epproc} tasks of the Science Analysis System (SAS) software, version 18.0.0, respectively for the MOS and the pn cameras.
The obtained event files were filtered for soft-proton contamination using the \texttt{espfilt} task, thus obtaining a screened exposure time of 108.2 ks for MOS1, 110.1 for MOS2 and 94.7 ks for the pn camera.
Images were background subtracted by adopting the double subtraction procedure described in \citet{2006A&A...453..567M}, retaining only events with \texttt{FLAG = 0} and \texttt{PATTERN $\leq$ 12}.
With this method we removed instrumental, particles and diffuse X-ray background from the images by using the Filter Wheel Closed (FWC)\footnote{\url{https://www.cosmos.esa.int/web/xmm-newton/filter-closed}} and the Blank Sky (BS)\footnote{\url{http://xmm-tools.cosmos.esa.int/external/xmm_calibration//background/bs_repository/blanksky_all.html}} files available at XMM ESAC web pages.
Count-rate images were obtained by mosaicking MOS 1 and MOS 2 maps and are vignetting corrected and adaptively smoothed (with the \texttt{asmooth} task) to a signal-to-noise ratio of 10.

We analyzed the EPIC-pn spectra extracted from the same regions adopted for the \nus\ spectral analysis.
To take into account the vignetting effect in the spectra, we added a ``weight'' column to the pn event file with the \texttt{evigweight} SAS command.
Spectra were extracted by using the \texttt{evselect} task, retaining only events with \texttt{FLAG = 0} and \texttt{PATTERN $\leq$ 4}.
For each spectrum, we produced the RMF and ARF files, with the \texttt{rmfgen} and the \texttt{arfgen} tasks, respectively. Spectra were rebinned so as to have at least 25 counts per bin.
For the background, we selected two extraction regions outside the shell: one, at south, for regions 1-5, and one, at north, for regions 6-11 and for the ``Hard Knot'' (see Sect. 3.3 and Fig. 1 for the region selection).
Spectral analysis of \nus\ and \xmm\ data was performed with the HEAsoft software XSPEC version 12.11.1 \citep{1996ASPC..101...17A}. Spectra from different cameras were fitted simultaneously. 

\subsection{\textit{Chandra} Observation}

To study in detail the morphology of the remnant, we also analyzed \emph{Chandra} observations performed between April 27, 2006 and August 3, 2006 for a total exposure time of 750 ks (Obs. ID: 6714, 6715, 6716, 6717, 6718, 7366; PI: S. Reynolds). Data were reprocessed with the CIAO v4.13 software using  CALDB 4.9.4. We reprocessed the data by using the \texttt{chandra\_repro} task. Mosaicked flux images were obtained by using the \texttt{merge\_obs} task.

\section{Results}\label{sec:results}

\subsection{Images}

\begin{figure*}[!t]
     \centering
     \includegraphics[width=\columnwidth]{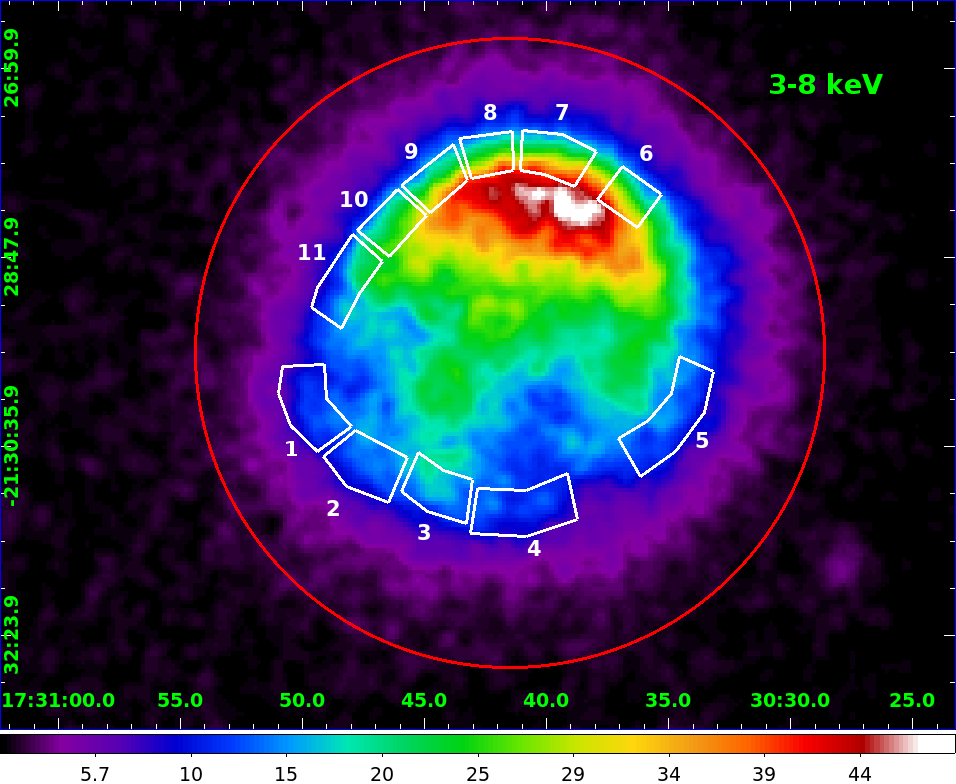}
     \includegraphics[width=\columnwidth]{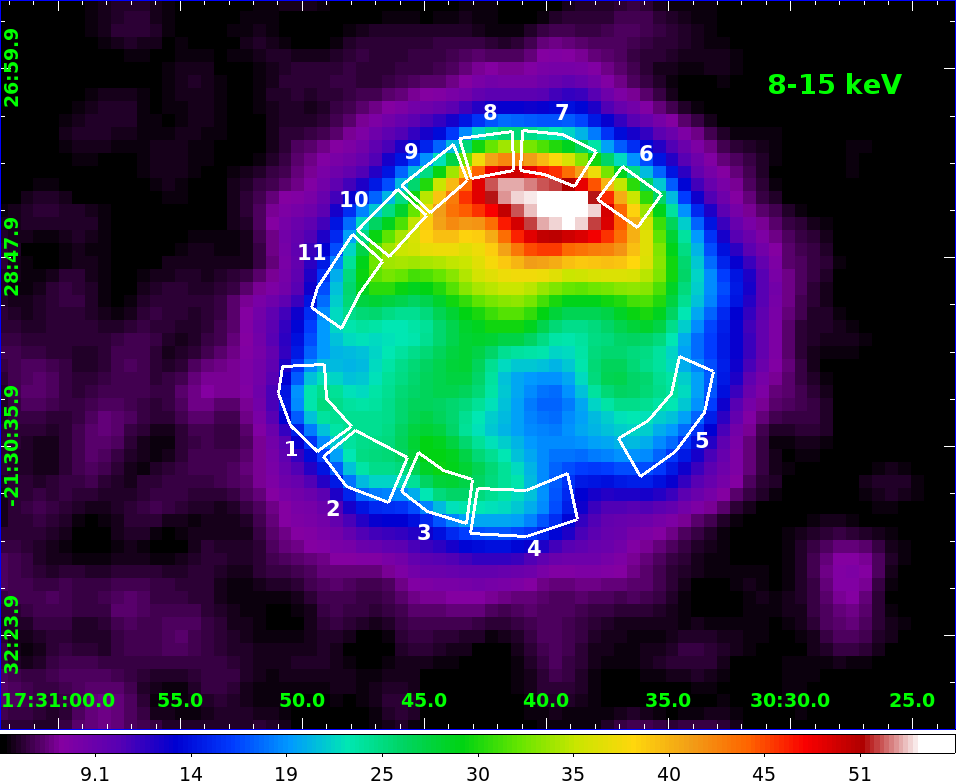}
     \includegraphics[width=\columnwidth]{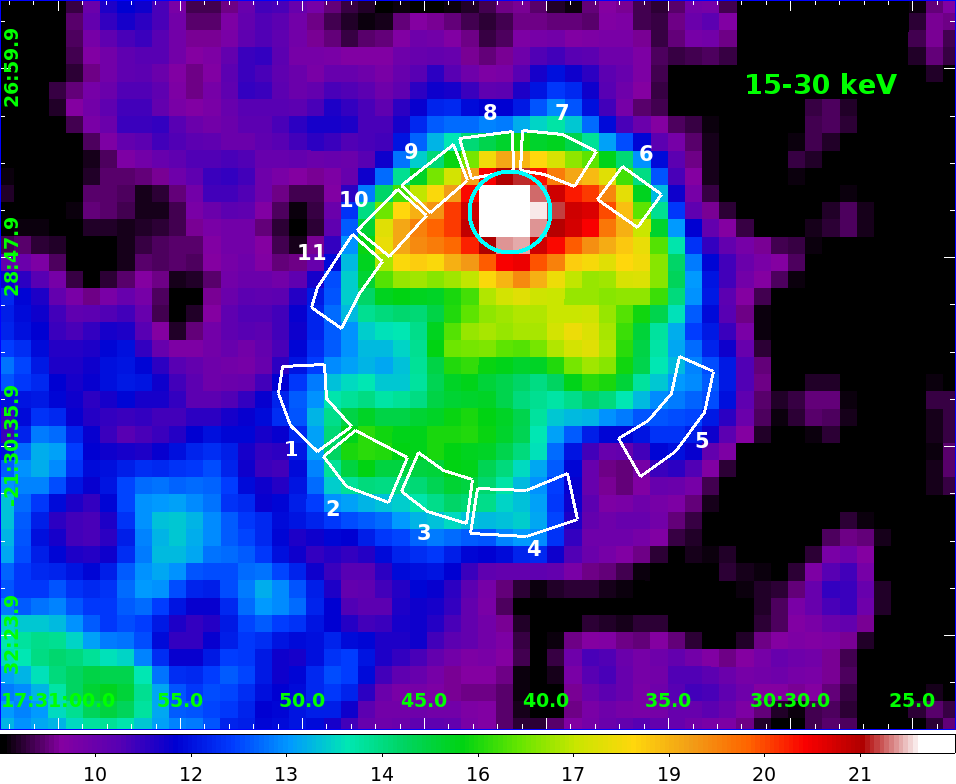}
     \includegraphics[width=\columnwidth]{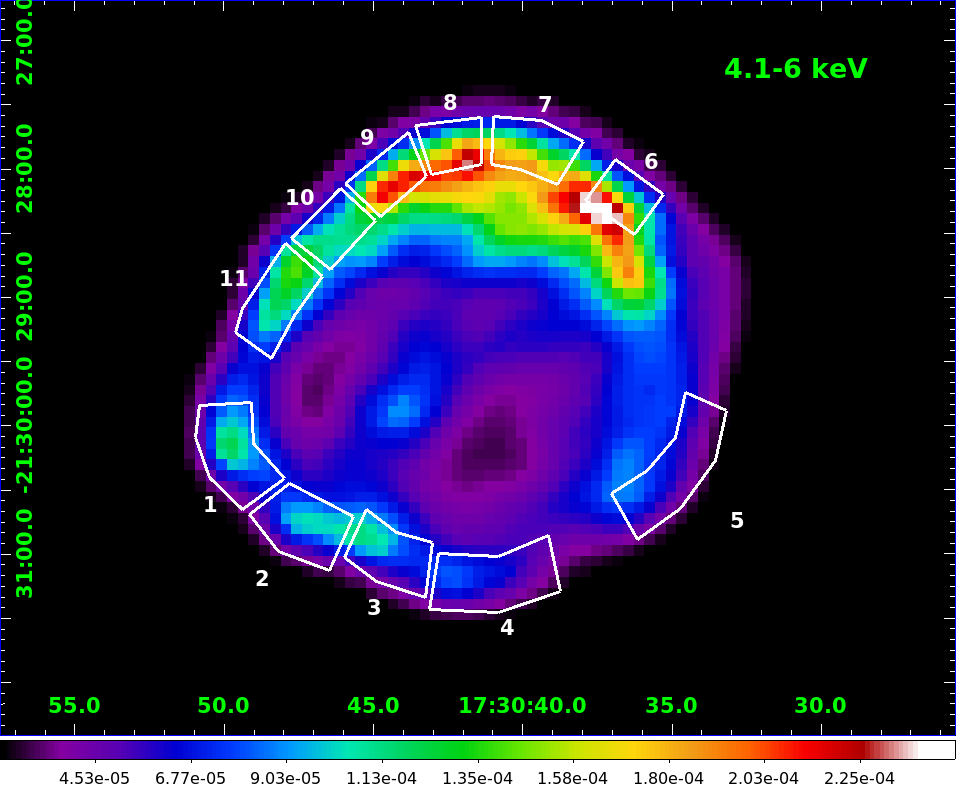}
     \caption{\textit{Upper-left panel:} \nus\ count image in the $3-8$ keV band in linear scale. The bin size is $2.5''$, and the image was smoothed through the convolution with a Gaussian with $\sigma = 7.5''$. The red circle marks the extraction region for the total spectrum (see Sect. \ref{subsec:totalspec}).
     \textit{Upper-right panel:} \nus\ count image in the $8-15$ keV band in linear scale. The bin size is $7.5''$, and the image was smoothed through the convolution with a Gaussian with $\sigma = 22.5''$.
     \textit{Bottom-left panel:}  \nus\ count image in the $15-30$ keV band in linear scale. The bin size is $10''$, and the image was smoothed through the convolution with a Gaussian with $\sigma = 30''$. The cyan circle marks the extraction region for the hard X-ray knot (see Sect. \ref{sec:hardknot}) Regions selected for the spatially resolved spectral analysis at the rim of the shell are indicated by white polygons.
     \textit{Bottom-right panel:} \xmm\ count-rate image in the $4.1-6$ keV band in linear scale. The bin size is $5''$ and the image was adaptively smoothed to a signal-to-noise ratio of 10.
     North is up and east is to the left.}
     \label{fig:3to8vs8to20}
 \end{figure*}

Fig. \ref{fig:3to8vs8to20} shows the \nus\ count image in the $3-8$ keV, $8-15$ keV, and $15-30$ keV bands, together with the \xmm\ count-rate image in the $4.1-6$ keV band.
The presence of source photons up to 30 keV confirms the detection of hard X-rays by \citet{2021PASJ...73..302N}.
Thanks to the angular resolution of \emph{NuSTAR}, we can also reveal the spatial distribution of the hard X-ray emission in Kepler's SNR.
Fig. \ref{fig:3to8vs8to20} shows that the morphology of the emission in the $8-15$ keV band is roughly similar to the soft X-ray emission, being brighter in the northern part of the shell, where the shock is interacting with the nitrogen-rich CSM. 
Similarly, in the $15-30$ keV band we observe a higher surface brightness in the north than in the south. However, some differences in the morphology of the hard X-ray emission with respect to the soft emission are visible, as, for example, the position of the peak in surface brightness in the $15-30$ keV band, which is located to the east with respect to  the peak in the $3-8$ keV map.
We also notice an enhancement in the surface brightness at southeast. 
The $4.1-6.0$ keV \xmm\ count-rate image is bright in the outermost regions of the remnant, where synchrotron filaments have been spotted (\citealt{2007ApJ...668L.135R}).
We point out that, because of the higher densities expected in the northern part of the shell, we also expect a larger contribution of thermal bremsstrahlung therein. The high-energy tail of thermal bremsstrahlung can, in principle, contribute to the $15-30$ keV emission in the northern part of the shell. However, as we show in Sect. \ref{Spectra}, the bulk of the hard X-ray emission of Kepler's SNR has likely a non-thermal origin.

\subsection{Total Spectrum}
\label{subsec:totalspec}

\begin{figure}
    \centering
    \includegraphics[width=\columnwidth]{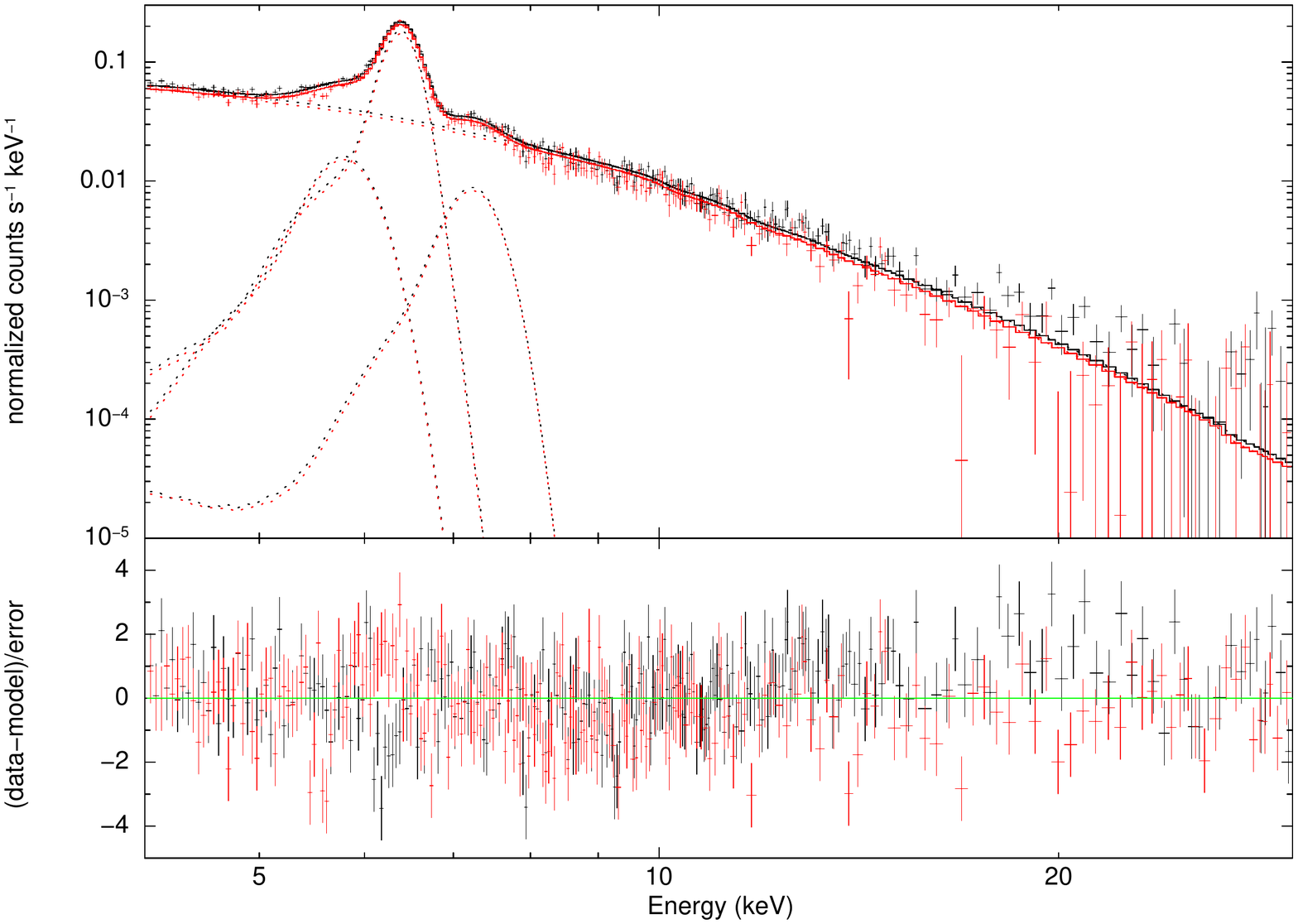}
    \caption{FPMA (black) and FPMB (red) total spectra of Kepler's SNR with the corresponding best-fit model and residual in the $4.1-30$ keV band.}
    \label{fig:tuttospec}
\end{figure}

We extracted the spectrum of the whole Kepler's SNR from \nus\ FPMA and FPMB using a circle with a radius of 3' and center coordinates $\alpha_{J2000} = 17^h30^m41.44^s$ and $\delta_{J2000} = -21$°$29' 27.7''$, shown in Fig. \ref{fig:3to8vs8to20}. 
We modeled the FPMA and FPMB global spectra in the $4.1-30$ keV band,
We excluded the $3-4.1$ keV band to avoid the strong contamination of thermal emission present in this energy band. 
We model the hard X-rays as non-thermal emission, by adopting a similar approach as \citet{2015ApJ...814..132L}.
However, here we describe the continuum as synchrotron emission from an electron energy distribution limited by radiative losses (hereafter loss-limited model, \citealt{2007A&A...465..695Z}), which has been shown to provide an accurate description of non-thermal X-ray emission in young SNRs (e.g., \citealt{za10,mc12,mbd13,2021ApJ...907..117T}). 
The spectrum of the loss-limited model is given by:
\begin{equation}
\label{eq:ziramodel}
    \frac{dN_X}{d\varepsilon} \propto \bigg(\frac{\varepsilon}{\varepsilon_0}\bigg)^{-2}\bigg[1+ 0.38\sqrt{\frac{\varepsilon}{\varepsilon_0}}\bigg]^{\frac{11}{4}} \exp \bigg( - \sqrt{\frac{\varepsilon}{\varepsilon_0}}\bigg),
\end{equation}
where $\varepsilon$ is the photon energy and $\varepsilon_0$ is the cutoff energy parameter.
We include interstellar absorption (\texttt{Tbabs} model in XSPEC), with a hydrogen column density fixed to $N_H = 6.4 \times 10^{21}$ cm$^{-2}$ (as in \citealt{2015ApJ...808...49K}).
We also include three ad-hoc Doppler-broadened Gaussian components to model the Fe K line (at $\sim 6.4$ keV, see \citealt{2013ApJ...766...44Y}), the Cr and Mn emission lines and the Ni emission lines.
Fig. \ref{fig:tuttospec} shows the total spectra (FPMA and FPMB) of Kepler's SNR with the corresponding best fit model and residuals.
The best-fit values, with error bars at 68\% confidence level, are reported in Tab. \ref{tab:totspec}.

\begin{table}
    \centering
    \caption{Best-ft values for Kepler's SNR \nus\ spectra.}
    \begin{tabular}{cc}
    \hline\hline
        Parameter & Value\\
        \hline
         Fe K center (keV)& 6.4018$_{-0.0014}^{+0.0016}$ \\
         $\sigma$ (keV)& 0.091$_{-0.003}^{+0.003}$\\
         Norm ($10^{-4}$ photons cm$^{-2}$ s$^{-1}$)& 2.96$_{-0.02}^{+0.02}$\\
         $\varepsilon_0$ (keV)&$0.640_{-0.013}^{+0.014}$\\
         norm ($10^{-3}$)&29.0$_{-1.7}^{+1.8}$\\
         $\chi^2/d.o.f.$& $1115.82/918$ \\
         \hline
    \end{tabular}
    \label{tab:totspec}
\end{table}
The fit provides a $\chi^2/d.o.f. =  1115.82 / 918$ and an average cutoff energy parameter $\varepsilon_0=0.640_{-0.013}^{+0.014}$ keV.
This value of $\varepsilon_0$ corresponds to a roll-off frequency $\nu_{r}=1.55_{-0.03}^{+0.03}\times 10^{17}$ Hz, which is in good agreement with the estimate obtained by \citet{2021PASJ...73..302N} ($~1\times10^{17}$ Hz) on the basis of the modeling of the broad band spectral energy distribution.

We fitted simultaneously the \nus\ FPMA and FPMB spectra (in the $4.1-30$ keV band), and the \emph{Suzaku} HXD-PIN spectrum (in the $15-30$ keV band),  with the loss-limited model. 
We allowed the normalization of the \emph{Suzaku} spectra to differ from that of the \nus\ spectra  within a $10\%$ to account for the characteristic cross-calibration factor between the two telescopes \citep{2017AJ....153....2M}. 
We found that the hard X-ray flux of Kepler's SNR in the $15-30$ keV band is $1.05_{-0.03}^{+0.04}\times 10^{-12}$ erg cm$^{-2}$ s$^{-1}$ ($F_X$ $1.15_{-0.24}^{+0.02}\times10^{-12}$ erg cm$^{-2}$ s$^{-1}$ for Suzaku HXD, taking into account the cross-calibration factor).
Though this value is lower than that reported by \citet{2021PASJ...73..302N} ($2.75_{-0.77-0.82}^{+0.78+0.81}\times10^{-12}$ erg cm$^{-2}$ s$^{-1}$), it is still consistent with it, considering the cross-calibration factor and the 90\%-statistical and systematic errors.

\subsection{Spatially resolved Spectral Analysis}
\label{Spectra}

We performed a spatially resolved spectral analysis by analyzing the spectra extracted from the eleven regions shown in Fig. \ref{fig:3to8vs8to20}.
We focus on the outer rim of the shell, by defining regions with similar photon counts in the $8-30$ keV band ($N_{8-30}\approx 800$) in order to investigate the relation between the shock velocity and the maximum energy of electrons accelerated at the shock front.

To ascertain the origin of the hard X-ray emission, we first focus on the \nus\ spectra in the $8-30$ keV band.
The emission in this band is characterized by a featureless continuum, which can be modelled with a power-law with spectral index $\Gamma\sim3$ in all the eleven regions considered. For example, in the southern part of the shell, we obtain $\Gamma=3.5_{-0.4}^{+0.5}$ in region 2 and $\Gamma=2.7_{-0.3}^{+0.3}$ in region 5; similarly, in the northern limb, $\Gamma=3.0_{-0.4}^{+0.4}$ in region 7 and $\Gamma=2.7_{-0.4}^{+0.5}$ in region 10. 
By modeling this relatively flat emission as a thermal bremsstrahlung, we derive quite high plasma temperatures (namely, $kT=5.8_{-1.2}^{+1.8}$ keV in region 2, $kT=9_{-2}^{+3}$ keV in region 5, $kT=7.8_{-1.8}^{+2.8}$ keV in region 7 and $kT=10_{-3}^{+5}$ keV in region 10).
We then consider the bulk of the hard X-ray emission to be non-thermal. However, we point out that, by including a thermal contribution to the hard X-ray spectra, our conclusions stay unaffected, as shown below. 

We then fitted the spectra in the $4.1-30$ keV band ($4.1-8$ keV for EPIC-pn and $4.1-30$ keV for \nus\ FPMA and FPMB) using the loss-limited model with an additional Gaussian component to take into account the Fe K line. We also added a further Gaussian component in regions 8-11 to model the Cr/Mn K line detected therein.
The spectra from all regions, with the corresponding best-fit model and residual, are shown in Appendix B, Fig. \ref{fig:1ziraonly}.
The best-fit values  for all regions (with the corresponding  $\chi^2 / d.o.f.$) are shown in Table \ref{tab:ecut} (error bars are at $68\%$ confidence level).

\begin{table*}
    \centering
    \caption{Best-fit values for spectra from regions labeled from 1 to 11. All errors are at the 68\% confidence level.} 

\begin{tabular}{ccccccc}

\hline\hline
 Region \# & Fe K center (keV) & $\sigma$ (keV) & norm ($10^{-6}$ photons cm$^{-2}$ s$^{-1}$)& $\varepsilon_0$ (keV) & norm ($10^{-3}$)& $\chi^2 / d.o.f.$ \\
 \hline
 1&$ 6.445_{-0.018}^{+0.019}$ & $<0.08$  & $ 0.99_{-0.12}^{+0.12}$  & $ 0.57_{-0.06}^{+0.07}$  & $ 0.65_{-0.18}^{+0.24}$  & 107.08/97  \\
 2&$ 6.38_{-0.04}^{+0.04}$ & $ 0.26_{-0.05}^{+0.06}$ & $ 0.66_{-0.08}^{+0.09}$  & $ 0.47_{-0.04}^{+0.05}$  & $ 0.48_{-0.13}^{+0.17}$  & 117.76/90 \\
 3&$6.427_{-0.016}^{+0.018}$ & $ 0.09_{-0.02}^{+0.02}$ & $ 1.21_{-0.11}^{+0.10}$  & $ 0.43_{-0.04}^{+0.04}$  & $ 1.1_{-0.3}^{+0.4}$  & 97.02/93 \\
 4&$ 6.395_{-0.014}^{+0.016}$ & $ 0.10_{-0.02}^{+0.02}$  & $ 2.27_{-0.18}^{+0.18}$  & $ 0.48_{-0.04}^{+0.05}$  & $1.3_{-0.3}^{+0.4}$  & 117.52/118 \\
 5&$ 6.410_{-0.008}^{+0.009}$ & $ 0.084_{-0.014}^{+0.014}$  & $ 3.87_{-0.19}^{+0.19}$  & $ 0.54_{-0.05}^{+0.06}$  & $ 0.9_{-0.2}^{+0.3}$  & 175.24/128 \\
 6&$ 6.441_{-0.006}^{+0.007}$ & $ 0.082_{-0.008}^{+0.008}  $ &$6.6_{-0.3}^{+0.3} $  & $ 0.50_{-0.04}^{+0.05}$  & $ 1.5_{-0.4}^{+0.5}$  & 168.34/121 \\
 7&$ 6.448_{-0.005}^{+0.005}$ & $ 0.083_{-0.007}^{+0.006}$  & $ 9.7_{-0.3}^{+0.3}$  & $ 0.59_{-0.05}^{+0.06}$  & $ 1.0_{-0.3}^{+0.3}$  & 189.15/154  \\
 8&$ 6.443_{-0.007}^{+0.006}$ & $ 0.091_{-0.008}^{+0.008}$  & $ 7.2_{-0.3}^{+0.3}$  & $ 0.67_{-0.07}^{+0.09}$  & $ 0.60_{-0.17}^{+0.24}$  & 123.26/111 \\
 9&$ 6.434_{-0.005}^{+0.005}$ & $ 0.067_{-0.007}^{+0.007}$  & $ 9.2_{-0.3}^{+0.3}$  & $ 0.64_{-0.07}^{+0.08}$  & $ 0.8_{-0.2}^{+0.3}$  & 158.83/131 \\
 10&$ 6.419_{-0.007}^{-0.006}$ & $ 0.078_{-0.009}^{+0.009}$  & $ 5.4_{-0.2}^{+0.2}$  & $ 0.59_{-0.06}^{+0.07}$  & $ 0.68_{-0.19}^{+0.26}$  & 108.60/109 \\
 11&$ 6.404_{-0.013}^{+0.012}$ & $0.09_{-0.02}^{+0.03}$  & $ 1.54_{-0.10}^{+0.10}$  & $ 0.59_{-0.06}^{+0.06}$  & $ 0.34_{-0.09}^{+0.12}$  & 121.37/98\\
\hline
 Hard Knot& $6.428_{-0.003}^{+0.004}$& $0.101_{-0.004}^{+0.004}$& 24.8$_{-0.6}^{+0.6}$& 0.70$_{-0.04}^{+0.04}$&$1.5_{-0.2}^{+0.2}$ & 466.54/340\\
 \hline
\end{tabular}
\label{tab:ecut}
\end{table*}

As a crosscheck, we verified that our assumption on the non-thermal origin of the hard X-ray emission is consistent with the broadband X-ray spectrum. To this end, we fit the spectra of all the regions in the $0.3-30$ keV ($0.3-8$ keV for EPIC-pn and $3-30$ keV for \nus\ FPMA and FPMB) by adding thermal components to the loss-limited model derived above. 

The soft X-ray spectra ($0.3-4.1$ keV) for the northern regions show prominent thermal emission features, as the Fe-L line complex in the $0.7-1.2$ keV band, and the Si, S, Ar, Ca K lines, respectively at $\sim1.86$, keV $\sim2.48$ keV, $\sim3.11$ keV and $\sim3.86$ keV. All the spectra are shown in Fig. \ref{fig:8bvrneibf}, with the corresponding best-fit model and residuals.

The spectra extracted from southern regions (regions $1-5$, characterized by a fainter thermal emission) are well described by two components of isothermal optically thin plasma in non-equilibrium of ionization (NEI) with non-solar abundances and Doppler broadening (\texttt{bvrnei} model in XSPEC), in addition to the absorbed loss-limited model (the parameters of the loss-limited model are fixed to the best-fit values shown in Tab. \ref{tab:ecut}).
We also included  two Gaussian lines to take into account missing Fe-L lines in the \texttt{bvrnei} model (as in \citealt{2015ApJ...808...49K}). 
The spectra extracted from northern regions (regions 6-11, characterized by a brighter thermal emission) were fitted by adding to the loss-limited model (with parameters fixed to the values in Tab. \ref{tab:ecut}) three thermal components.
We found a degeneracy between the Fe abundance, the normalization and the temperature ($kT_h$) of the hottest component of the plasma.
\citet{2021PASJ...73..302N} found  $kT_h=3.74_{-0.03}^{+0.12}$ keV so we decided to set an upper limit of 4 keV for $kT_h$.
The values of the best-fit parameters are listed in Tables \ref{tab:broadsouth} and \ref{tab:braodnorth}.
This model provides a good description of the spectra of all regions ($1<\chi^2 / d.o.f.<1.4$).
We conclude that the modeling of the synchrotron emission adopted in the analysis of the ``hard'' spectra is consistent with the broadband X-ray spectra and provides a robust description of the non-thermal emission in Kepler's SNR.

\subsection{Hard X-ray knot}
\label{sec:hardknot}

Lastly, we analyzed the spectrum of the knot with the brightest hard X-ray emission, which we spotted in the 15-30 keV band map of Kepler's SNR.
In particular, we extracted the EPIC-pn, FPMA and FPMB spectra from the circular region indicated by the cyan circle in the lower left panel of Fig. \ref{fig:3to8vs8to20}.

We modelled the spectra in the the $4.1-30$ keV band ($4.1-8$ keV for EPIC-pn and $4.1-30$ keV for \nus\ FPMA and FPMB) by adopting the loss-limited model (with the additional Gaussian component) described in detail in Sect. \ref{Spectra}.
\begin{figure}
    \centering
    \includegraphics[width=\columnwidth]{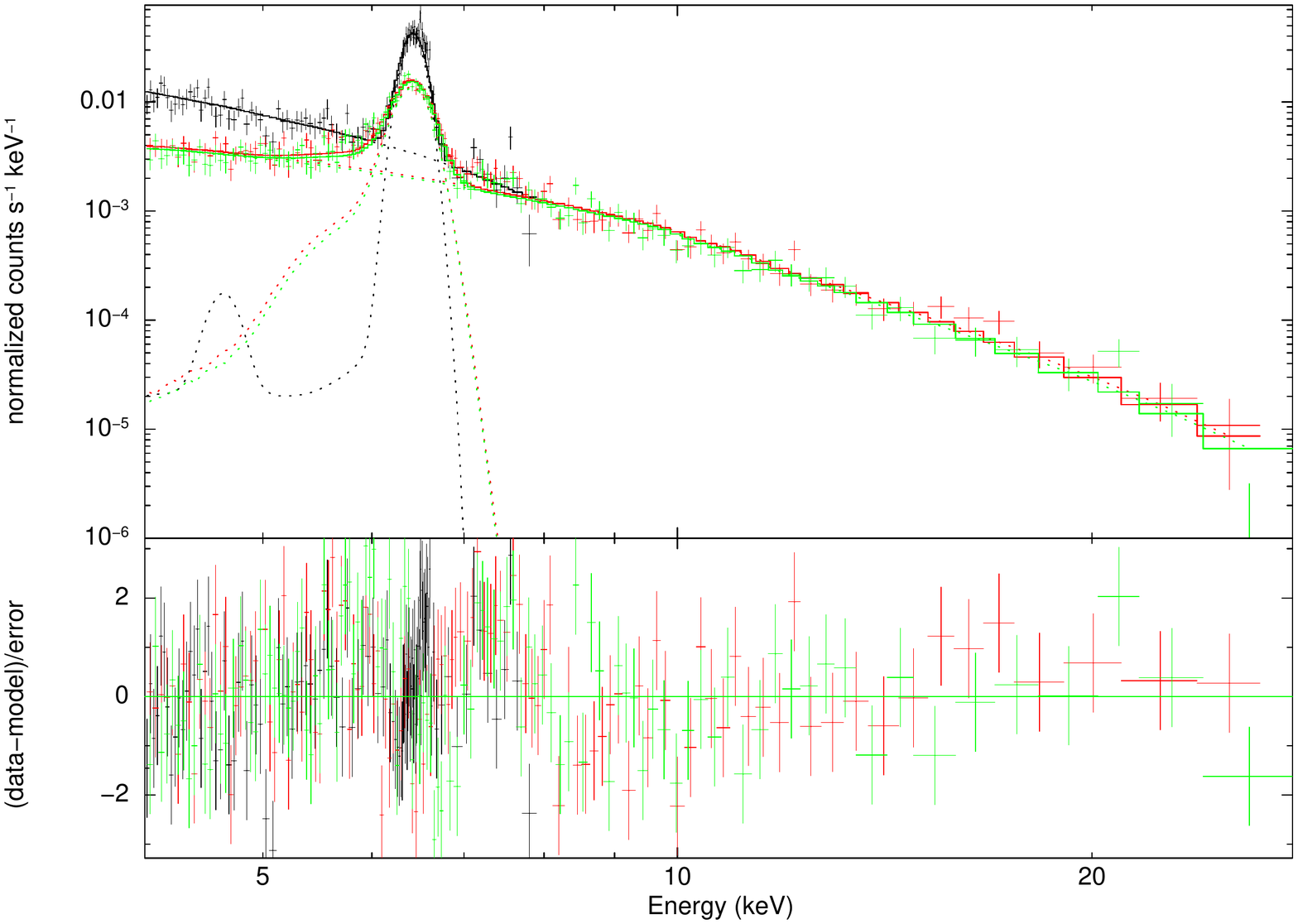}
    \caption{EPIC-pn (black), FPMA (red) and FPMB (green) spectra of the hardest knot in Kepler's SNR (cyan circle in Fig. \ref{fig:3to8vs8to20}) with the corresponding best-fit model and residual in the $4.1-30$ keV band.}
    \label{fig:hardknot}
\end{figure}
Spectra from the hard knot, with the corresponding best-fit model and residual, are shown in Fig. \ref{fig:hardknot}, while the best fit values (with errors at the 68\% confidence level) are shown in Tab. \ref{tab:ecut}.

\section{Discussion}
\label{sec:conclusions}

\subsection{Different regimes of particle acceleration in Kepler's SNR}
\label{subsec:ecut}
The cutoff energy parameter $\varepsilon_0$ is a crucial factor in order to characterize the acceleration mechanism in SNRs, because it is determined by the balance between acceleration and cooling in the synchrotron emission process.
\citet{2007A&A...465..695Z} derived a relation between $\varepsilon_0$, the Bohm factor and the shock velocity,

\begin{equation} 
\label{eq:eovsvsh}
    \varepsilon_0=\frac{1.6}{\eta}\bigg( \frac{v_{sh}}{4000\text{ km s}^{-1}}\bigg)^2 \text{keV}.
\end{equation}

\begin{figure*}
    \centering
    \includegraphics[width=0.49\textwidth]{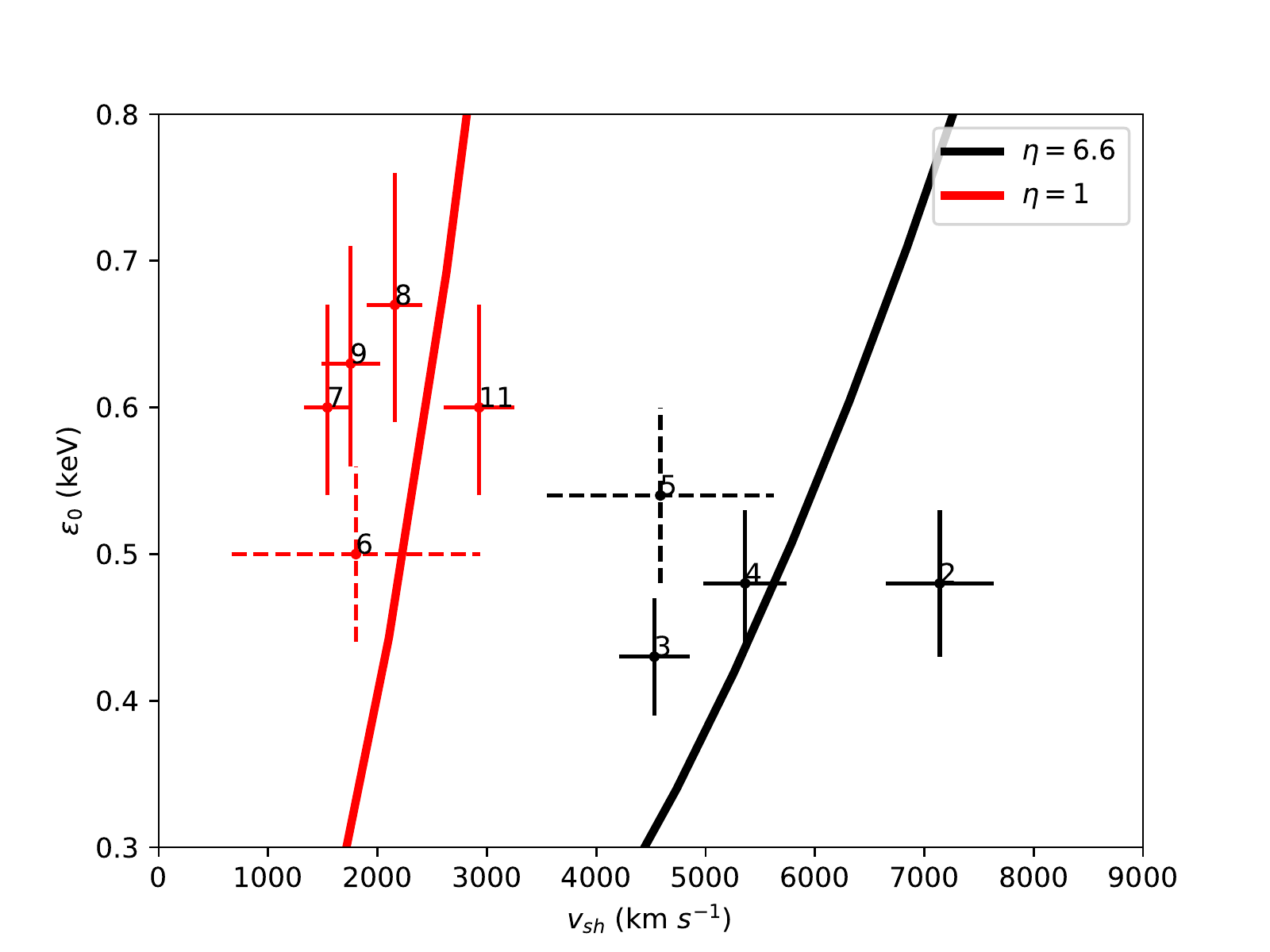}  
    \includegraphics[width=0.49\textwidth]{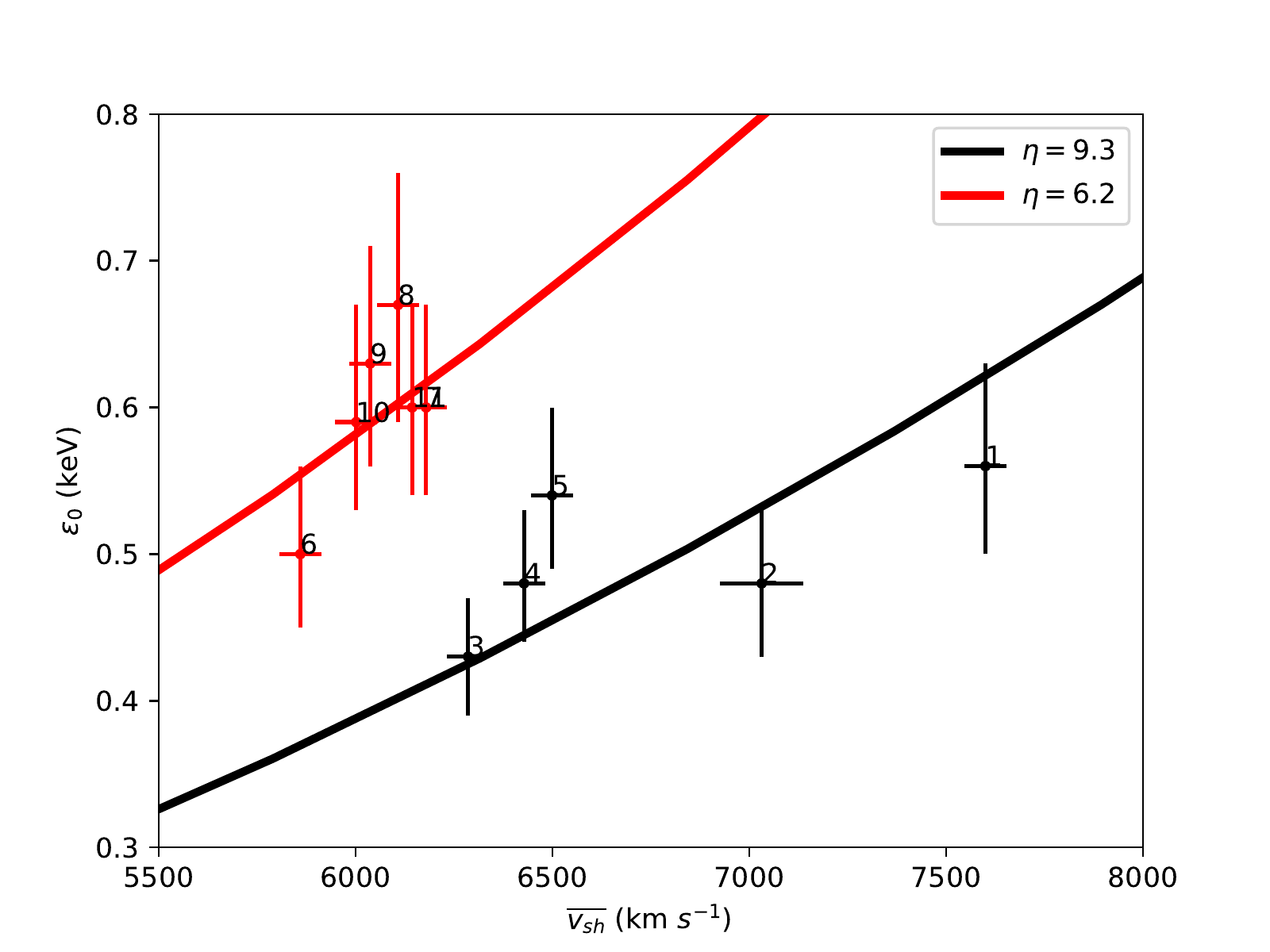}
    \caption{\textit{left panel:} Synchrotron cutoff energy vs. current shock velocity derived from \citet{2022ApJ...926...84C}, (solid crosses) and \citealt{Katsuda_2008} (dashed crosses, see Appendix \ref{app:pm} for details).
    Red crosses mark northern regions (6-9/11) and the red solid curve is the corresponding best-fit curve obtained from Eq. \ref{eq:eovsvsh}. Black crosses mark southern regions (2-5) and the black solid curve is the corresponding best-fit curve obtained from Eq. \ref{eq:eovsvsh}.
    \textit{right panel:} Synchrotron cutoff energy vs. average shock speed for Kepler's SNR. Red crosses mark regions 6-11, and the red solid curve is the corresponding best-fit curve obtained from Eq. \ref{eq:eovsvsh}. Black crosses mark regions 1-5, and the black solid curve is the corresponding best-fit curve obtained from Eq. \ref{eq:eovsvsh}.}
    \label{fig:Ecutvsr}
\end{figure*}

By adopting the same approach as \citet{2021ApJ...907..117T}, we show in the left panel of Fig. \ref{fig:Ecutvsr} the values of $\varepsilon_0$ obtained from the spectral fittings of the regions with a proper motion measurement available either in \citet{Katsuda_2008} or in \citet{2022ApJ...926...84C}, as a function of the corresponding $v_{sh}$.
We mark with different colors data points derived for southern regions (black) and northern regions (red).  In northern regions we obtain, on average, lower shock velocities and higher $\varepsilon_0$ values than in southern regions ($\varepsilon_0^S=0.48\pm 0.02$ keV, $\varepsilon_0^N=0.60\pm 0.03$ keV, for regions 2-5 and regions 6-9/11, respectively).
Moreover, the figure shows that southern and northern regions identify two distinct clusters.
This suggests the presence of two different regimes of electron acceleration in the same SNR.
If we describe each of the two clusters with Eq. \ref{eq:eovsvsh}, we can derive the corresponding best fit values of the Bohm diffusing factor, obtaining  $\eta=6.6\pm 1.6$ in the north and reaching the Bohm limit in the south.
These values are similar to those derived by \citet{2021ApJ...907..117T}, who find $\eta\sim0.3-4$. However, \citet{2021ApJ...907..117T} adopted a distance of 4 kpc (instead of 5 kpc) to derive the shock velocity from the proper motion measurements, thus obtaining lower velocities than those reported in the left panel of Fig. \ref{fig:Ecutvsr}, which is in line with their lower values of $\eta$.

Taking into account these results, we can estimate the acceleration time of electrons (\citealt{2001RPPh...64..429M,2020pesr.book.....V}) as
\begin{equation}
    t_{acc}\approx 756\frac{\eta}{\delta}\sqrt{\frac{\varepsilon_0}{1~\rm{keV}}}\bigg( \frac{v_{sh}}{5000~\rm{km~s}^{-1}}\bigg)^{-2}\bigg( \frac{B}{10~\mu\rm{G} }\bigg)^{-\frac{3}{2}}~\rm{yr}\label{eq:tacc}
\end{equation}
where $\delta$ accounts for the energy dependence of the diffusion coefficient, and typically ranges between 0.3 and 0.7 (see \citealt{2007ARNPS..57..285S}).
Considering $\delta=0.5$ and taking the values we found in the northern shell ($v_{sh}=1800$ km s$^{-1}$, $\eta=1$ and $\varepsilon_0=0.64$ keV), we derive $t_{acc} \approx 300\cdot\bigg(\frac{B}{100 \mu\text{G}} \bigg)^{-\frac{3}{2}}$yr.
Such a high value of the acceleration time may suggest that the electrons started to accelerate before the interaction of the shock with the CSM, which produced a deceleration of the northern shock front. 
\citet{1999A&A...351..330A} suggested that the shock velocity should be larger than 2000~km~s$^{-1}$ to emit synchrotron X-rays (see also \citealt{2008ApJ...689..231V}).
Thus the existence of synchrotron X-rays implies that the deceleration should happen recently.
Similar situation happens in a super-bubble with synchrotron X-rays, 30~Dor~C \citep{2004ApJ...602..257B}, where the supernova shock just hit the shell of the super-bubble and emit synchrotron X-rays \citep{2009PASJ...61S.175Y} although \citet{2020ApJ...893..144L} suggests the super-bubble itself accelerate electrons.

In this framework, the current shock velocity may not be representative of the shock conditions over the whole acceleration process.
We then explore an alternative scenario, by studying the relationship between the synchrotron cutoff energy and the average shock velocity, $\overline{v_{sh}}$ in all the regions selected for our spatially resolved spectral analysis. We derived $\overline{v_{sh}}$ for each region as $\overline{v_{sh}}=r_{sh}/t_{age}$, where $r_{sh}$ is the radius of the shock.
We estimated $r_{sh}$ for regions 1-11 by measuring the distance between the shock front and the center of the remnant (whose position was carefully derived by \citealt{2017ApJ...845..167S}).The measuring procedure was performed detecting the edge at the azimuthal center of each region on the \textit{Chandra} flux image in the $4.1-6$ keV band, to exploit the high spatial resolution of the \textit{Chandra} mirrors (we associated to the angular distance an error of 1.5'').

Fig. \ref{fig:Ecutvsr} (\textit{right panel}) shows the values of $\varepsilon_0$ obtained from the spectral fittings of the eleven regions analyzed, as a function of the corresponding $\overline{v_{sh}}$. Again, in northern regions we obtain, on average, lower shock velocities and higher $\varepsilon_0$ values than in southern regions ($\varepsilon_0^S=0.48\pm 0.02$ keV, $\varepsilon_0^N=0.59\pm 0.02$ keV, for regions 1-5 and regions 6-11, respectively). Moreover, also in this plot, southern and northern regions clearly identify two distinct clusters.
Therefore, by adopting the average shock velocities, we recover the presence of two different regimes of electron acceleration in the Kepler's SNR.
Each of the two clusters can be well described by Eq. \ref{eq:eovsvsh} with a specific value of $\eta$. 
We then derive the corresponding best fit values of the Bohm diffusing factor, obtaining  $\eta=9.3\pm 0.4$ in the north and $\eta=6.2\pm 0.2$ in the south, with a null hypothesis probability of $\sim90$\%. 
These values of $\eta$ are higher than those derived with the current shock velocities. 
This is because the deceleration of the shock front makes the current velocities systematically lower than the average velocities (see Eq. \ref{eq:eovsvsh} for the dependence of $\eta$ on $v_{sh}$).
Therefore, the values of the Bohm factors should be taken with some caution.

The two scenarios considered above, namely synchrotron emission originating from  i) electrons accelerated in the current shock conditions (i. e., freshly accelerated electrons in a high magnetic field) and ii) electrons accelerated well before the interaction with the dense CSM at north (i. e., longer acceleration times, possibly associated with a lower magnetic field, as in \citealt{2021PASJ...73..302N}), may be considered as two limiting cases, bracketing the actual evolution of the system.

Nevertheless, regardless of the shock velocity adopted (current velocity vs. average velocity),  our conclusions do not change, since both the plots shown in Fig. \ref{fig:Ecutvsr} point toward a more efficient (i.e. closer to the Bohm limit) electron acceleration in the north than in the south.

On the other hand, a scenario in which $\varepsilon_0$ does not depend on the shock velocity is also possible. We tested this possibility by fitting the data points of Fig. \ref{fig:Ecutvsr} with a constant  $\varepsilon_0$, and obtained a null hypothesis probability of $\sim15\%$, which is well below the value obtained in the loss-limited case, but still statistically acceptable. However, a framework where $\varepsilon_0$ does not depend on  $v_{sh}$ would indicate that the maximum electron energy is not limited by radiative losses and, as explained in Sect. \ref{sec:intro}, this would imply a magnetic field lower than $30~\mu$G (assuming $\varepsilon_0$=0.5 keV), which is at odds with the observations (\citealt{2005A&A...433..229V}, \citealt{2006A&A...453..387P}, \citealt{2012A&A...545A..47R}, \citealt{2021ApJ...917...55R}).

The cutoff energies discussed above were obtained by assuming that the X-ray continuum above 4 keV is ascribed to synchrotron radiation.
We then checked how a possible contamination of thermal emission in the hard X-ray spectra of northern regions (where thermal emission is the highest) affects our results. 
We found that if we model the hard continuum of regions 6-11 with a combination of synchrotron radiation and thermal bremsstrahlung, the value of $\varepsilon_0$ systematically increases with the contribution of thermal emission. 
The values of $\varepsilon_0$ for regions 6-11 shown in Table \ref{tab:ecut} and Fig. \ref{fig:Ecutvsr} should then be considered as lower limits. This means that the acceleration efficiency in the north may be even higher (and the Bohm diffusing factor lower) than that derived by neglecting the contribution of thermal emission to the hard X-ray continuum.
We conclude that the evidence of more efficient acceleration in the northern part of Kepler's SNR is solid.
Southern regions, where the thermal emission has a lower surface brightness, are less affected by thermal contamination and the synchrotron component is better constrained.
We note that \citet{2021ApJ...907..117T} finds small-scale variations in  $\varepsilon_0$, with local peaks reaching values of the order of 1.5 keV (higher than those reported in our Table \ref{tab:ecut}).
This may be due to the absence of \nus\ data for Kepler's SNR in \citet{2021ApJ...907..117T} and to the different size of the extraction regions (our regions being significantly larger than theirs).

We note that in the southern part of Kepler's SNR, where the ambient density is similar to that observed in Tycho's SNR \citep{2007ApJ...662..998B}, we recover similar results as those obtained for Tycho's SNR by \citet{2015ApJ...814..132L}, who found that the cutoff energy increases with the shock velocity (i.e.,  where the ambient density is low). 
At odds with Tycho's SNR, however, Kepler's SNR is interacting with a much denser environment in the north (4-7 times denser, \citealt{2007ApJ...662..998B,Katsuda_2008}), where we register a different regime of particle acceleration.

The presence of two different acceleration regimes and of a higher acceleration efficiency in regions 6-11 might be explained by considering the turbulent magnetic field generated in the interaction between the shock front and the dense CSM in the north.
\citet{2012ApJ...744...71I} modelling RX J1713.7-3946 SNR with a 3D magneto-hydrodynamic (MHD) simulation, show that a shock wave that sweeps a cloudy medium generates an amplified magnetic field, as a result of the dynamo action induced by the turbulent shock-cloud interaction (See also \citealt{2008ApJ...678..274O}).
An amplified magnetic field may lead toward a more efficient acceleration process and to a lower $\eta$ Bohm factor.
This interpretation is in line with the findings obtained for RX J1713.7-3946 by \citet{2015ApJ...799..175S}, who observed that the synchrotron photon index is anti-correlated with the X-ray intensity.
Indeed, we found the highest value of $\varepsilon_0$ in the ``Hard Knot'' region, which we identified in the northern part of the shell as a bright feature in the $15-30$ keV map (see Fig. \ref{fig:3to8vs8to20} and Table \ref{tab:ecut}). In general, our findings show that in Kepler's SNR the $\varepsilon_0$ parameter is high in the region where the shock interacts with high density CSM, thus indicating a similar scenario as that proposed for RX J1713.7-3946. 

\subsection{Spectral Energy Distribution}

Several SNRs are known to emit $\gamma$-rays up to TeV energy (e.g.\citealt{2013Sci...339..807A}, \citealt{2018A&A...612A...5H}, \citealt{2012A&A...541A..13A}).
\citet{2021ApJ...908...22X} reported a likely detection (with $\sim 4\sigma$ significance) of $\gamma$-ray emission in the $0.2-500$ GeV band from the region of Kepler's SNR by analyzing Fermi Large Area Telescope (LAT) data.
Using the same Fermi LAT data, \citet{2022arXiv220105567A} confirmed this detection up to $6.0\sigma$.
Moreover, \citet{2021arXiv210711582P} reported the detection of VHE emission from Kepler's SNR based on a deep observation of the High Energy Stereoscopic System (H.E.S.S.).

In this section, we model the Spectral Energy Distribution (SED) of Kepler's SNR for the non-thermal emission, using radio \citep{2002ApJ...580..914D}, X-ray (\nus\ FPMA and FPMB in the $8-30$ keV band from this work and HXD-PIN from \citealt{2021PASJ...73..302N}), GeV \citep{2021ApJ...908...22X} and TeV \citep{2021arXiv210711582P} data, and give some constraints on the particle energy distribution, and on the ambient density and magnetic field.

\begin{figure*}
    \centering
    \includegraphics[width=\columnwidth]{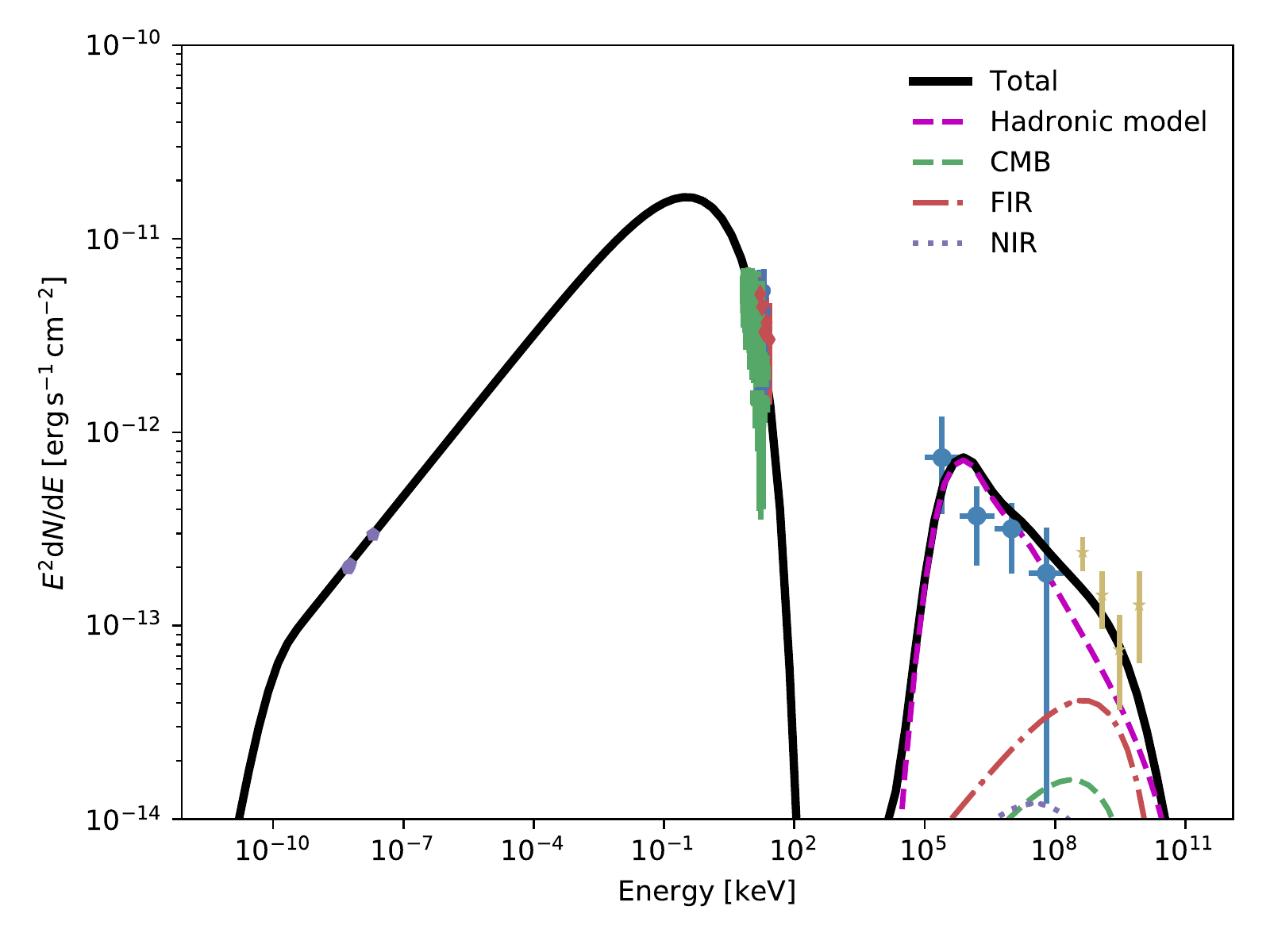}
    \includegraphics[width=\columnwidth]{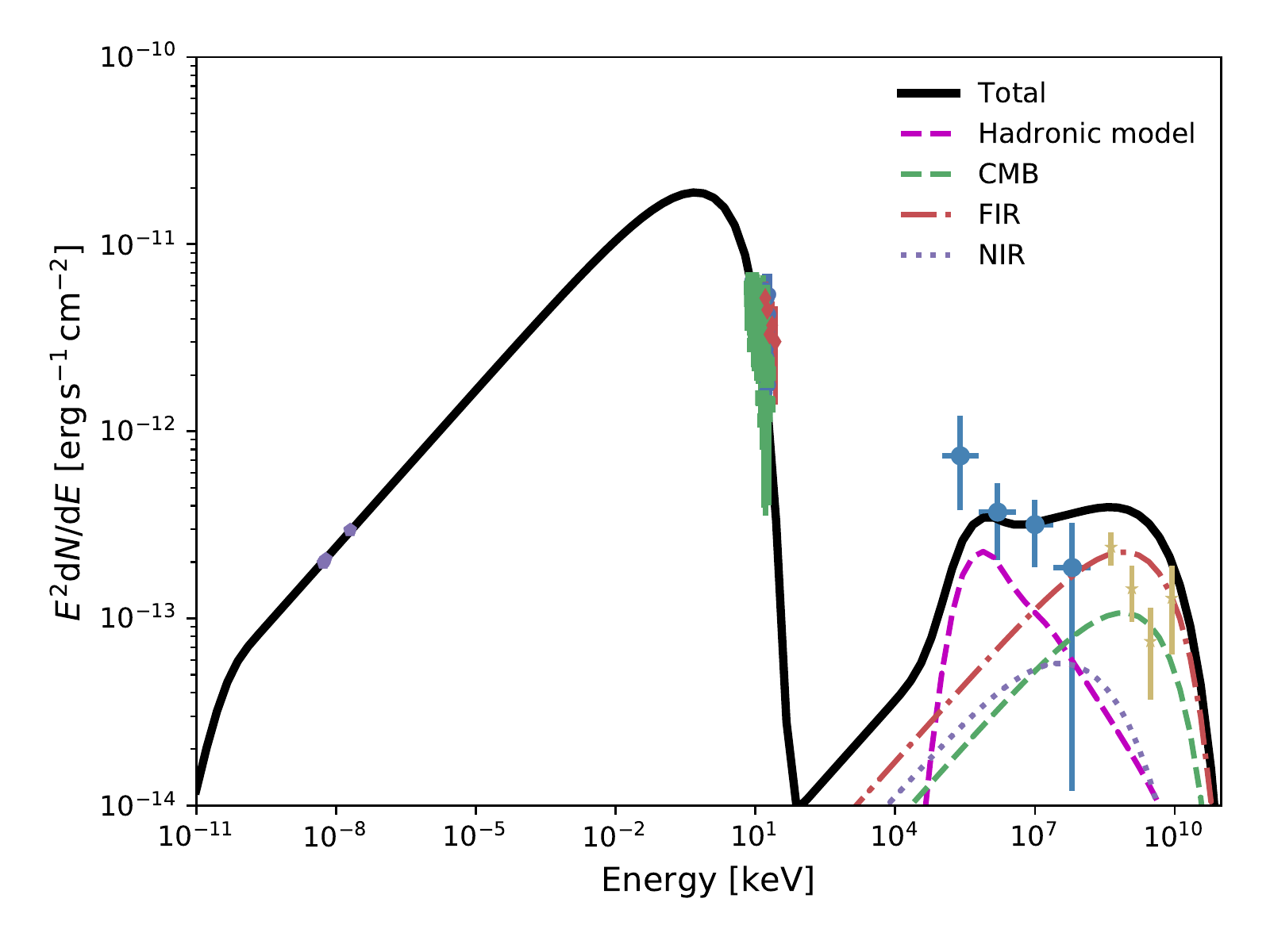}
    \caption{\textit{Left panel:} Spectral Energy Distribution of Kepler's SNR.
    Radio data (violet) are from \citet{2002ApJ...580..914D}, NuSTAR (FPMA and FPMB, blue and green respectively) X-ray data are extracted from the whole remnant, HXD-PIN X-ray data (dark red) are from \citet{2021PASJ...73..302N}, GeV $\gamma$-ray data (steel blue) are from \citet{2022arXiv220105567A} and TeV  $\gamma$-ray data (yellow) are from \citet{2021arXiv210711582P}. The black curve indicates our combined lepto-hadronic model, including contributions from $\pi^0$ decay (dashed magenta curve), and Inverse Compton emission from the cosmic microwave background (green dashed curve), far-infrared (red, dash-dotted curve) and near-infrared emission (purple dotted curve).
    In the lepto-hadronic model, the magnetic field is 100 $\mu$G and the post shock density is 20 cm$^{-3}$.
    \textit{Right panel:} Same as left panel, but with a  magnetic field of 40 $\mu$G and a post shock density of 5 cm$^{-3}$.
    }
    \label{fig:SED}
\end{figure*}

We use the radiative code \texttt{naima} (version 0.9.1, \citealt{naima}) to model the SED.
We considered a lepto-hadronic one-zone stationary model to describe the multi-band emission spectrum.
In this model, the synchrotron and Inverse Compton (IC) emission are assumed to stem from the same electron distribution, which is described by a power-law with an exponential cutoff.
We considered the same seed photon field as \citet{2021PASJ...73..302N} for the IC emission:
the cosmic microwave background radiation (CMBR), a far-infrared (FIR) component ($T=29.5$ K and $u_{FIR}=1.08$ eV cm$^{-3}$), and a near-infrared (NIR) component ($T=1800$ K and $u_{NIR}=2.25$ eV cm$^{-3}$).
As for the hadrons, we assumed a power law energy distribution with an exponential cutoff. 
The model that best reproduces the observed data is shown in Fig. \ref{fig:SED} (left panel).

Our model gives for the leptonic part a spectral index $\alpha=2.44$ and a cutoff energy $E_{cut}=16$ TeV.
The electrons emit synchrotron radiation in a magnetic field of 100 $\mu$G.
For the hadronic part we assumed the same spectral index and an energy cutoff of 500 TeV. 
The total kinetic energy of protons ($W_p$), interacting with a post shock medium with density of 20 cm$^{-3}$ (consistent with density value of the shock-CSM interacting region, \citealt{2021ApJ...915...42K}), was set to be 15 times the electrons total kinetic energy ($W_p=4.2\times10^{48}$ erg).
As a comparison we also modeled the SED using the value for the magnetic field proposed by \citealt{2021PASJ...73..302N} (B=40 $\mu$G, see Fig. \ref{fig:SED} right panel), with $\alpha=2.44$ and $E_{cut}=35$ TeV for the leptonic part.
For the hadronic part of the model we adopted the same assumption as before but with $W_p=5.4\times10^{48}$ erg (4 times the electrons total kinetic energy) and a post shock medium density of 5 cm$^{-3}$.
This last case gives a poor fit compared with the case where the magnetic field is higher.

\section{Conclusions}
\label{conc}

We analyzed an archive \nus\ observation of Kepler's SNR. 
We detected hard X-ray emission up to $\sim30$ keV, mainly stemming from the northern part of the remnant, where the shock front is interacting with dense circumstellar material.
We verified that the bulk of the hard X-ray emission is non-thermal and performed a spatially resolved spectral analysis focusing on the outer rim of the shell by combining \nus\ and \xmm\ spectra.
We adopted the loss-limited synchrotron emission model by \citet{2007A&A...465..695Z} to determine the cutoff photon energy parameter in 11 regions. 
We identified two different acceleration regimes in the northern and southern limbs of Kepler's SNR. 
In particular, we found lower Bohm factors (i.e. more efficient electron acceleration) in the northern part of the shell than in the south. 
We suggest that the interaction of the shock front with the high density CSM at north generates an amplified, possibly turbulent, magnetic field, which facilitates the particle acceleration process.

An alternative scenario of constant cutoff energy across the shell of Kepler's SNR is disfavoured by our analysis.
This would imply that the maximum electron energy is not limited by radiative losses (which would require that the magnetic field is much smaller than that observed), though it cannot be statistically excluded.

We produced the spectral energy distribution including also \nus\ X-ray data. 
We were able to reproduce all the data with a lepto-hadronic model with a magnetic field of 100 $\mu$G, a medium density of 20 cm$^{-3}$, electron energy $W_e=2.7\times10^{47}$ erg and proton energy $W_p=4.2\times10^{48}$ erg.
The density we found in modeling the SED is consistent with that derived in the northern region by \citet{2021ApJ...915...42K}, suggesting a scenario in which the majority of the hadronic emission originates in the northern part of the remnant.
The bright non-thermal hard X-ray emission that we detected in the northern half of Kepler's SNR strongly suggests that this region is also site of leptonic emission.


\begin{acknowledgments}
\indent \small The authors are truly thankful to the anonymous referee for their helpful comments and uplifting suggestions.
This work was partially supported by ``FFR 2021 Marco Miceli" of the University of Palermo in the framework of the STARS project.
M.M., F.B., S.O., and G.P. acknowledge financial contribution from the PRIN INAF 2019 grant ``From massive stars to supernovae and supernova remnants: driving mass, energy and cosmic rays in our Galaxy'’ and the INAF mainstream program ``Understanding particle acceleration in galactic sources in the CTA era''.
A.B. S.K. and Y.T. acknowledge financial support from the Grants-in-Aid for Scientific Research from the Japanese Ministry of Education, Culture, Sports, Science and
Technology (MEXT) of Japan, No. 19K03908 (A.B.) JP21H01121 (S.K. and Y.T.).
Y.T. and S.K .are deeply appreciative of the Observational Astrophysics Institute at Saitama University for supporting the research fund.
\end{acknowledgments}

\software{NuSTARDAS (v2.0.0), HEAsoft (v6.28; HEASARC 2014), SAS (v18.0.0; Gabriel et al. 2004), XSPEC (v12.11.1; Arnaud 1996), CIAO (v4.13; Fruscione et al. 2006), naima (v0.9.1, Zabalza 2015)}
%


\appendix
\section{Best-fit values for broadband spectra}

\begin{table*}[h!]
    \caption{Best-fit values for broadband spectra of southern regions (1-5).
    Solar abundances from \citet{1989GeCoA..53..197A}.
    $\tau$ lower limit set to be $5\times10^8$ cm$^{-3}$ s.
    Abundances upper limit is set to be 1000 times the solar one.
    Velocity upper limit set to be $1\times10^4$ m s$^{-1}$.
    Abundances consistent with their solar values were fixed to 1.}
    \resizebox{\textwidth}{!}{
    \begin{tabular}{ccccccc}
    \hline\hline
         Component & Parameter &\#1 &\#2 &\#3 &\#4 &\#5 \\
         \hline
         \texttt{TBabs} & $n_H$ (10$^{22}$ cm$^{-2}$) & \multicolumn{5}{c}{0.64(fixed)}\\
         \hline
         \multirow{2}{*}{\texttt{gauss$_{1}$}}&Fe L+O K (keV)&\multicolumn{5}{c}{0.708(fixed)}\\
         &Norm ($10^{-4}$ photons cm$^{-2}$ s$^{-1}$)&$2.8_{-0.7}^{+0.6}$&$1.7_{-0.7}^{+0.7}$&$4.3_{-0.9}^{+0.9}$&$11.0_{-1.2}^{+1.3}$&11.6$_{-2.1}^{+1.3}$\\
         \multirow{2}{*}{\texttt{gauss$_{2}$}}&Fe L+Ne K (keV)&\multicolumn{5}{c}{1.227(fixed)}\\
         &Norm ($10^{-4}$ photons cm$^{-2}$ s$^{-1}$)&$0.053_{-0.014}^{+0.017}$&$0.026_{-0.014}^{+0.014}$&$0.051_{-0.016}^{+0.016}$&$0.14_{-0.02}^{+0.02}$&0.15$_{-0.02}^{+0.03}$\\
         \hline
         \multirow{12}{*}{\texttt{brvnei$_1$}}& kT$_1$ (keV)&$0.63_{-0.05}^{+0.09}$&$0.65_{-0.09}^{0.11}$&$0.54_{-0.06}^{+0.06}$&0.68$_{-0.07}^{+0.03}$&0.721$_{-0.073}^{+0.007}$\\
         &O&1 (fixed)&$1.5_{0.9}^{+1.0}$&2.2$_{-0.6}^{+0.3}$&0.31$_{-0.17}^{+0.18}$&1 (fixed)\\
         &Ne&1 (fixed)&$1.3_{-0.9}^{+1.1}$&1 (fixed)&$<0.5$&1 (fixed)\\
         &Mg&1 (fixed)&$6_{-3}^{+4}$&4.80485$_{-1.4}^{+1.7}$&0.9$_{-0.5}^{+0.5}$&4.3$_{-0.7}^{+0.8}$\\
         &Si&\multicolumn{3}{c}{1 (fixed)}&9.0$_{-1.8}^{+1.9}$&\multicolumn{1}{c}{1 (fixed)}\\
         &Fe&$19_{-3}^{+5}$&$81_{-15}^{+18}$&51$_{-7}^{+12}$&50$_{-5}^{+4}$&107$_{-5}^{+4}$\\
         &$\tau_1$ (10$^{9}$ cm$^{-3}$ s)&$<0.6$&$1.05_{-0.13}^{+0.14}$&$<0.62$&0.83$_{-0.12}^{+0.07}$&1.20$_{-0.03}^{+0.04}$\\
         &Velocity (10$^{4}$ km s$^{-1}$)&0 &$0.7_{-0.5}^{+0.3}$&0&0&0.64$_{-0.10}^{+0.07}$\\
         &EM$_1$ (10$^{56}$ cm$^{-3}$)&$2.9_{-0.8}^{+0.4}$&$0.62_{-0.14}^{+0.20}$&$2.4_{-0.8}^{+0.7}$&4.6$_{-0.4}^{+1.1}$&4.16$_{-0.27}^{+0.10}$\\
         \multirow{13}{*}{\texttt{brvnei$_2$}}& kT$_2$ (keV)&$>3.88$&$>3.89$&$>3.93$&$>3.91$&$>3.95$\\
         &O&1 (fixed)&$<2.5$&$<1.7$&5$_{-2}^{+3}$&1 (fixed)\\
         &Ne&$3.7_{-1.4}^{+5.3}$&$2.4_{-1.0}^{+1.2}$&$4.2_{-0.8}^{1.3}$&5.2$_{-1.3}^{+2.1}$&$23_{-3}^{+10}$\\
         &Mg&$5_{-2}^{+7}$&$1.2_{-0.8}^{+0.9}$&$1.2_{-0.5}^{+0.7}$&2.3$_{-0.7}^{1.0}$&14.0$_{-3.0}^{+1.3}$\\
         &Si&$45_{-1.7}^{+64}$&$20_{-4}^{+6}$&$30_{-6}^{+10}$&37$_{-8}^{+14}$&170$_{-40}^{+720}$\\
         &S&$50_{-20}^{+80}$&$23_{-6}^{+9}$&$38_{-9}^{14}$&48$_{-11}^{+19}$&$300_{-40}^{+700}$\\
         &Ar&$50_{-40}^{+100}$&$26_{-19}^{+24}$&$40_{-20}^{+30}$&36$_{-17}^{+25}$&$180_{-50}^{+40}$\\
         &Ca&\multicolumn{3}{c}{1 (fixed)}&$24_{-22}^{+30}$&$110_{-80}^{+70}$\\
         &Fe&$27_{-2}^{+3}$&$13.8_{-1.1}^{1.2}$&$21.5_{-1.2}^{+1.3}$&26.8$_{-1.3}^{+1.1}$&$43.5_{-1.7}^{+0.7}$\\
         &Ni&  \multicolumn{5}{c}{=Fe}\\
         &$\tau_2$ (10$^{9}$cm$^{-3}$ s)&$4.51_{0.17}^{+2.3}$&$5.3_{-0.2}^{+0.3}$&$4.65_{-0.12}^{+0.14}$&5.12$_{-0.13}^{+0.13}$&7.14$_{-0.07}^{+0.10}$\\
         &Velocity (10$^{4}$ km s$^{-1}$)&$0.58_{-0.09}^{+0.09}$&$0.738_{-0.10}^{+0.09}$&$0.42_{-0.07}^{+0.06}$&0.42$_{-0.05}^{+0.05}$&0.57$_{-0.03}^{0.02}$\\
         &EM$_2$ (10$^{56}$ cm$^{-3}$)&$1.959_{-0.004}^{+0.003}$&$0.050_{-0.004}^{+0.004}$&$0.074_{-0.004}^{+0.003}$&0.113$_{-0.006}^{+0.007}$&0.095$_{-0.007}^{+0.006}$\\
         \hline
         &$\chi^2/d.o.f.$&458.98/456&403.14/355&423.15/443&608.79/573&824.00/626\\
         \hline
    \end{tabular}
}
\label{tab:broadsouth}
\end{table*}

\begin{table*}[]
    \caption{
    Best-fit values for broadband spectra of northern regions (6-11).
    Solar abundances from \citet{1989GeCoA..53..197A}.
    $\tau$ lower limit set to be $5\times10^8$ cm$^{-3}$ s.
    Abundances upper limit is set to be 1000 times the solar one.
    Velocity upper limit set to be $1\times10^4$ m s$^{-1}$.
    Abundances consistent with their solar values were fixed to 1. }
    \resizebox{\columnwidth}{!}{
    \begin{tabular}{cccccccc}
    \hline\hline
         Component & Parameter &\#6 &\#7 &\#8 &\#9 &\#10 &\#11\\
         \hline
         \texttt{TBabs} & $n_H$ (10$^{22}$ cm$^{-2}$) & \multicolumn{6}{c}{0.64(fixed)}\\
         \hline
         \multirow{2}{*}{\texttt{gauss$_{1}$}}&Fe L+O K (keV)&\multicolumn{6}{c}{0.708(fixed)}\\
         &Norm ($10^{-4}$ photons cm$^{-2}$ s$^{-1}$)&18.5$_{-1.8}^{+1.7}$&$17.16_{-0.15}^{+0.15}$&$8.6_{-1.3}^{+1.3}$&25.2$_{-1.5}^{+1.7}$&22.1$_{-1.4}^{+1.2}$&$5.9_{-0.7}^{+1.2}$\\
         \multirow{2}{*}{\texttt{gauss$_{2}$}}&Fe L+Ne K (keV)&\multicolumn{6}{c}{1.227(fixed)}\\
         &Norm ($10^{-4}$ photons cm$^{-2}$ s$^{-1}$)&0.83$_{-0.05}^{+0.07}$&$1.413_{-0.06}^{+0.06}$&$1.02_{-0.05}^{+0.05}$&1.18$_{-0.06}^{+0.06}$&0.584099$_{-0.04}^{+0.04}$&$0.23_{-0.02}^{+0.02}$\\
         \multirow{2}{*}{\texttt{gauss$_{3}$}}&Cr K+Mn K (keV)&\multicolumn{6}{c}{5.6 (fixed)}\\
         &Norm ($10^{-4}$ photons cm$^{-2}$ s$^{-1}$)&/&/&$0.003_{-0.002}^{+0.002}$&0.006$_{-0.002}^{+0.002}$&0.0024$_{-0.0016}^{+0.0017}$&$0.0029_{-0.0009}^{+0.0009}$\\
         \hline
         \multirow{12}{*}{\texttt{brvnei$_1$}}& kT$_1$ (keV)&0.409$_{-0.017}^{+0.011}$&$0.431_{-0.013}^{+0.006}$&$0.442_{-0.013}^{+0.015}$&$0.430_{-0.008}^{+0.007}$&0.341519$_{-0.014}^{+0.013}$&$0.389980_{-0.02}^{+0.04}$\\
         &C&2.0$_{-0.9}^{+0.9}$&7.8$_{-1.8}^{+3.0}$ &$11_{-3}^{+6}$ &7.0$_{-1.7}^{+2.3}$ &31 $_{-6}^{+6}$&$<7$\\
         &N&\multicolumn{6}{c}{0 (fixed)}\\
         &O&$0.41_{-0.04}^{+0.05}$&$0.58_{-0.09}^{+0.16}$&$0.90_{-0.2}^{+0.4}$&$0.68_{-0.10}^{+0.14}$&1.40$_{-0.15}^{+0.15}$&0.34$_{-0.09}^{+0.12}$\\
         &Ne&4.1$_{-0.4}^{+0.4}$&$7.1_{-1.0}^{+1.8}$&$6.6_{-1.3}^{+2.6}$&6.1$_{-0.8}^{+1.1}$&12.8$_{-1.7}^{+1.7}$&2.9$_{-0.5}^{+0.9}$\\
         &Mg&$5.0_{-0.6}^{+0.7}$&$12.3_{-1.7}^{3.1}$&$10_{-2}^{+4}$&$11.6_{-1.3}^{+2.1}$&29$_{-4}^{+4}$&1 (fixed)\\
         &Fe&$51.4_{-5.7}^{1.4}$&$93_{-15}^{+27}$&$100_{-20}^{+40}$&$109_{-16}^{+22}$&275$_{-9}^{+12}$&70$_{-15}^{+24}$\\
         &Ni&\multicolumn{6}{c}{=Fe}\\
         &$\tau_1$ (10$^{9}$cm$^{-3}$ s)&$1.23_{-0.07}^{+0.06} $&$1.41_{-0.04}^{+0.05}$&$1.50_{-0.05}^{+0.06} $&$1.30_{-0.03}^{+0.03}$&0.993$_{-0.04}^{+0.05}$&1.08$_{-0.07}^{+0.07}$\\
         &Velocity (10$^{4}$ km s$^{-1}$)&$>0.74$&$0.55_{-0.05}^{+0.05}$&0 (fixed) &0.60$_{-0.05}^{+0.05}$&$0.36_{-0.13}^{+0.11}$&0.65$_{-0.16}^{+0.17}$\\
         &EM$_1$ (10$^{56}$ cm$^{-3}$)&$32_{-3}^{+4}$&$16_{-4}^{+3}$&$8_{-2}^{+2}$&$18_{-3}^{+3}$&$7.5_{-0.3}^{+0.3}$&$7.4_{-1.8}^{+1.8}$\\
         \multirow{13}{*}{\texttt{brvnei$_2$}}& kT$_2$ (keV)&0.97$_{-0.05}^{+0.06}$&$1.32_{-0.06}^{+0.06}$&1.43$_{-0.06}^{0.07}$&1.23$_{-0.04}^{+0.06}$&1.15$_{-0.03}^{+0.06}$&1.50$_{-0.16}^{+0.20}$\\
         &O&42$_{-9}^{+10}$&20$_{-6}^{+7}$&27$_{-8}^{+10}$&33$_{-9}^{+10}$&11.2175$_{-2}^{+2}$&30$_{-26}^{+26}$\\
         &Ne&46$_{-9}^{+10}$&\multicolumn{2}{c}{1 (fixed)}&40$_{-10}^{+9}$&9.00606$_{-2}^{+3}$&22$_{-21}^{+16}$\\
         &Mg&58$_{-12}^{+11}$&61$_{-10}^{+8}$&72$_{-17}^{+14}$&69$_{-15}^{+12}$&18.4$_{-1.8}^{+2.2}$&46$_{-40}^{+16}$\\
         &Si&180$_{-60}^{+30}$&230$_{-40}^{+30}$&270$_{-70}^{+50}$&$280_{-70}^{+50}$&114$_{-9}^{+11}$&300$_{-270}^{+100}$\\
         &S&290$_{-90}^{+60}$&380$_{-60}^{+50}$&470$_{-90}^{+80}$&480$_{-90}^{+90}$&183$_{-16}^{+18}$&450$_{-390}^{+160}$\\
         &Ar&350$_{-90}^{+100}$&370$_{-70}^{+80}$&520$_{-140}^{+110}$&520$_{-140}^{+130}$&240$_{-30}^{+30}$&250$_{-220}^{+240}$\\
         &Ca&$>600$&$>800$&$>700$&$>800$&360$_{-80}^{+90}$&$>100$\\
         &Fe&28$_{-6}^{+6}$&69$_{-12}^{+9}$&90$_{-23}^{+17}$&70$_{-18}^{+11}$&24.8$_{-0.8}^{+2.4}$&1 (fixed)\\
         &Ni&\multicolumn{6}{c}{=Fe}\\
         &$\tau_2$ (10$^{9}$cm$^{-3}$ s)&57$_{-6}^{+6}$&$43_{-3}^{+3}$&$36_{-2}^{+2}$&$39.9_{-2.8}^{+1.8}$&38.0$_{-4.4}^{+1.1}$&32$_{-5}^{+6}$\\
         &Velocity (10$^{4}$ km s$^{-1}$)&$0.394_{-0.017}^{+0.017}$&$0.533_{-0.014}^{+0.014}$&$0.438_{-0.015}^{+0.016}$&$0.39_{-0.05}^{+0.05}$&$0.480_{-0.014}^{+0.014}$&0.48$_{-0.04}^{+0.03}$\\
         &EM$_2$ (10$^{56}$ cm$^{-3}$)&0.121$_{-0.010}^{+0.003}$&$0.074_{-0.008}^{+0.016}$&$0.044_{-0.007}^{+0.015}$&$0.068_{-0.011}^{+0.022}$&$0.140_{-0.012}^{+0.012}$&$0.011_{-0.003}^{+0.079}$\\
         \multirow{13}{*}{\texttt{brvnei$_3$}}& kT$_3$ (keV)&$>3.91$&$>3.98$&$>3.97$&$>3.95$&$>3.96$&$>3.88$\\
         &Ne&$40_{-30}^{+30}$&\multicolumn{5}{c}{1 (fixed)}\\
         &Si&\multicolumn{2}{c}{1 (fixed)}&$<50$&\multicolumn{2}{c}{1 (fixed)}&28$_{-17}^{+146}$\\
         &S&$134_{-133}^{+134}$&\multicolumn{5}{c}{1 (fixed)}\\
         &Ar&$600_{-400}^{+400}$&$<500$&1 (fixed)&$<500$&\multicolumn{2}{c}{1 (fixed)}\\
         &Ca&\multicolumn{5}{c}{1 (fixed)}&190$_{-110}^{+470}$\\
         &Fe&$>900$&$>800$&$>800$&$>900$&$>970$&190$_{-80}^{+720}$\\
         &Ni&\multicolumn{6}{c}{=Fe}\\
         &$\tau_3$ (10$^{9}$cm$^{-3}$ s)&$6.49_{0.22}^{0.18}$&$6.78_{-0.11}^{+0.12}$&$6.32_{-0.11}^{+0.12}$&6.50$_{-0.11}^{+0.09}$&$6.31_{-0.13}^{+0.13}$&6.49$_{-0.21}^{+0.19}$\\
         &Velocity (10$^{4}$ km s$^{-1}$)&0.59$_{-0.06}^{+0.06}$&$0.53_{-0.04}^{+0.04}$&$0.59_{-0.05}^{+0.05}$&$0.43_{-0.05}^{+0.05}$&1$0.51_{-0.06}^{+0.06}$&$>0.9$\\
         &EM$_3$ (10$^{56}$ cm$^{-3}$)&$0.0100_{-0.0004}^{+0.0006}$&$0.0134_{-0.0002}^{+0.0027}$&$0.0099_{-0.0002}^{+0.0032}$&$0.0137_{-0.0002}^{+0.0024}$&0.0078$_{-0.0002}^{+0.0002}$&$0.014_{-0.011}^{+0.008}$\\
         \hline
         &$\chi^2/d.o.f.$&834.12/662&927.60/713&748.64/644&860.19/681&7748.25/628&545.85/494\\
         \hline

    \end{tabular}
}
\label{tab:braodnorth}
\end{table*}

\clearpage
\section{Spectra}

\begin{figure}[h!]
    \centering
    \includegraphics[width=0.3\columnwidth]{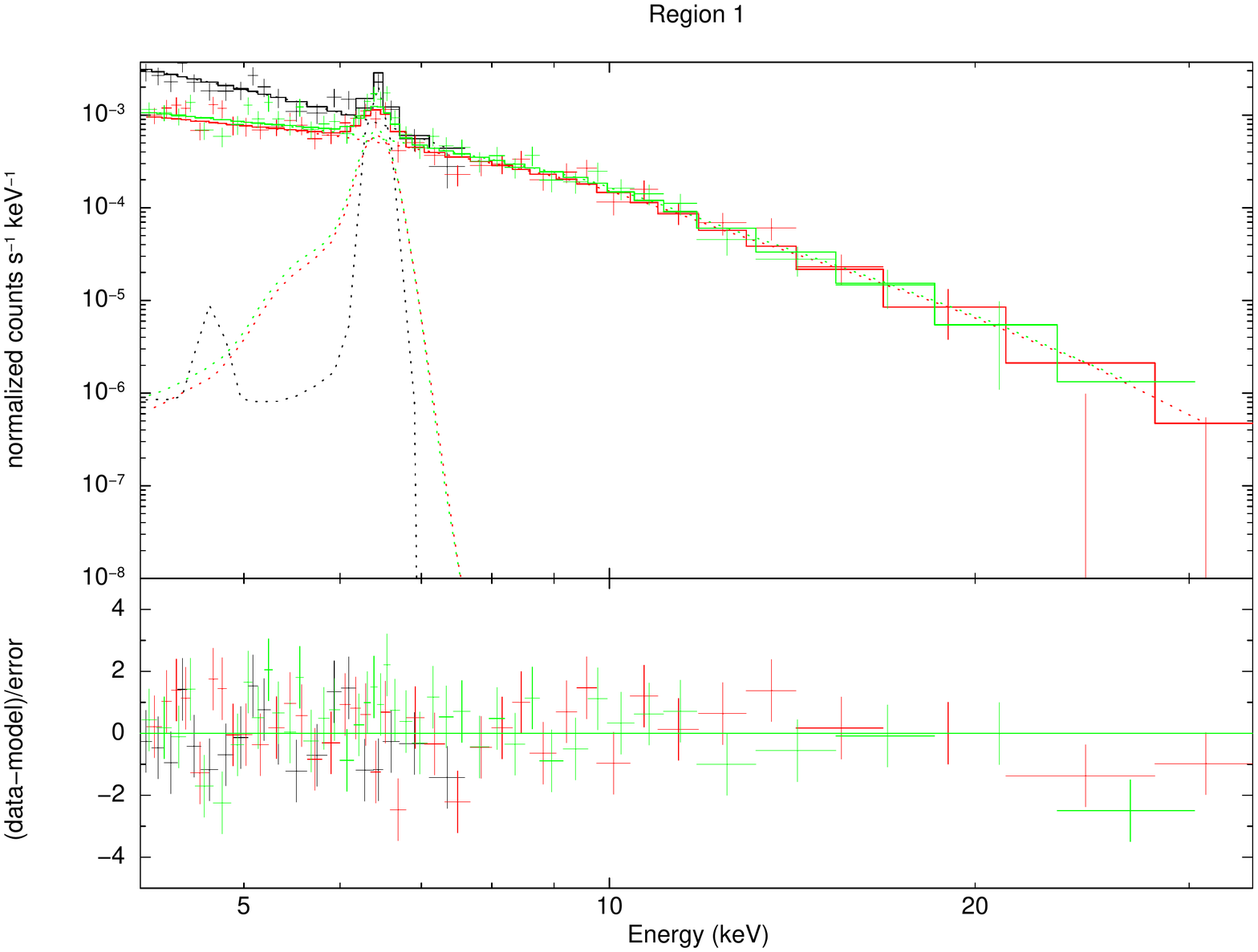}
    \includegraphics[width=0.3\columnwidth]{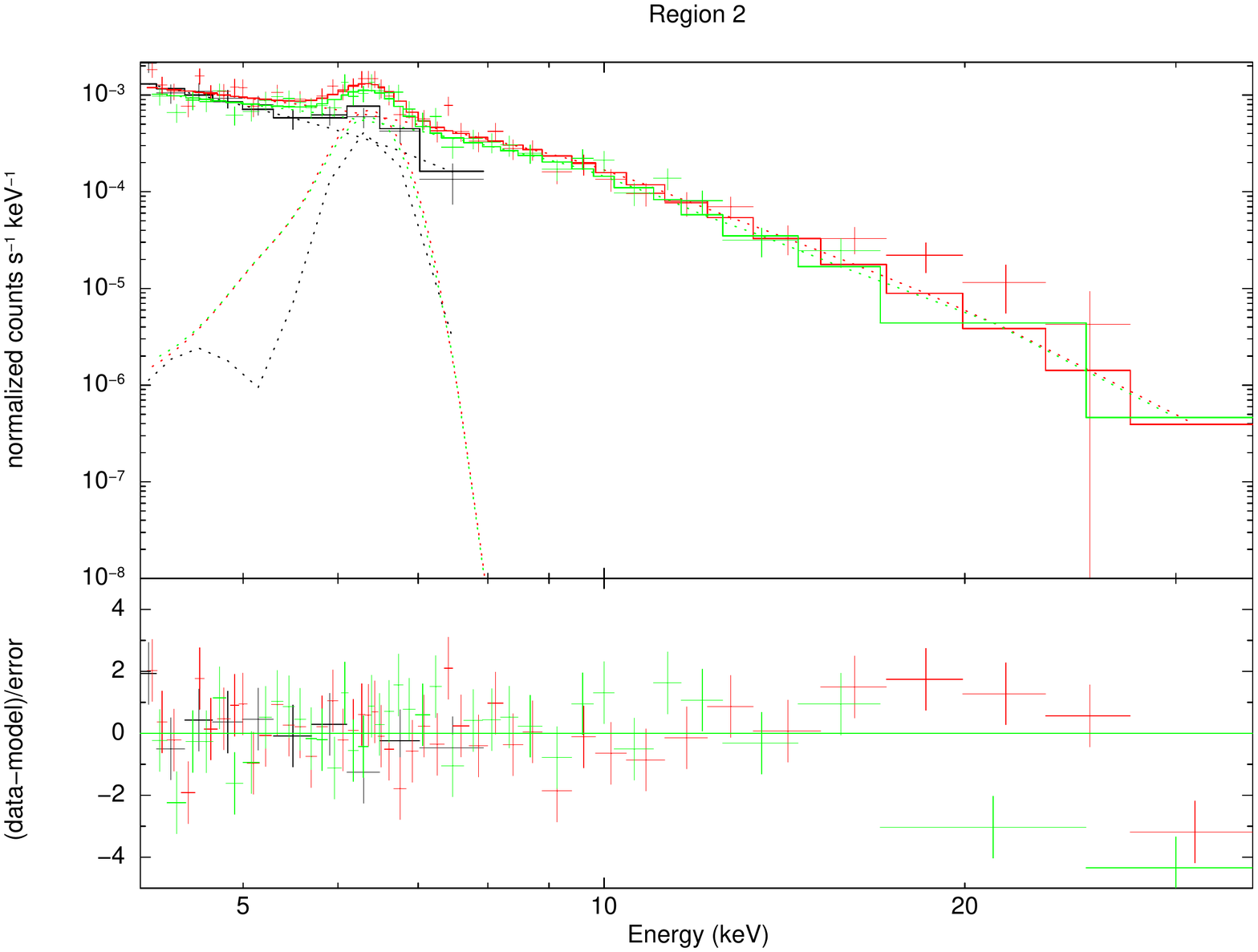}
    \includegraphics[width=0.3\columnwidth]{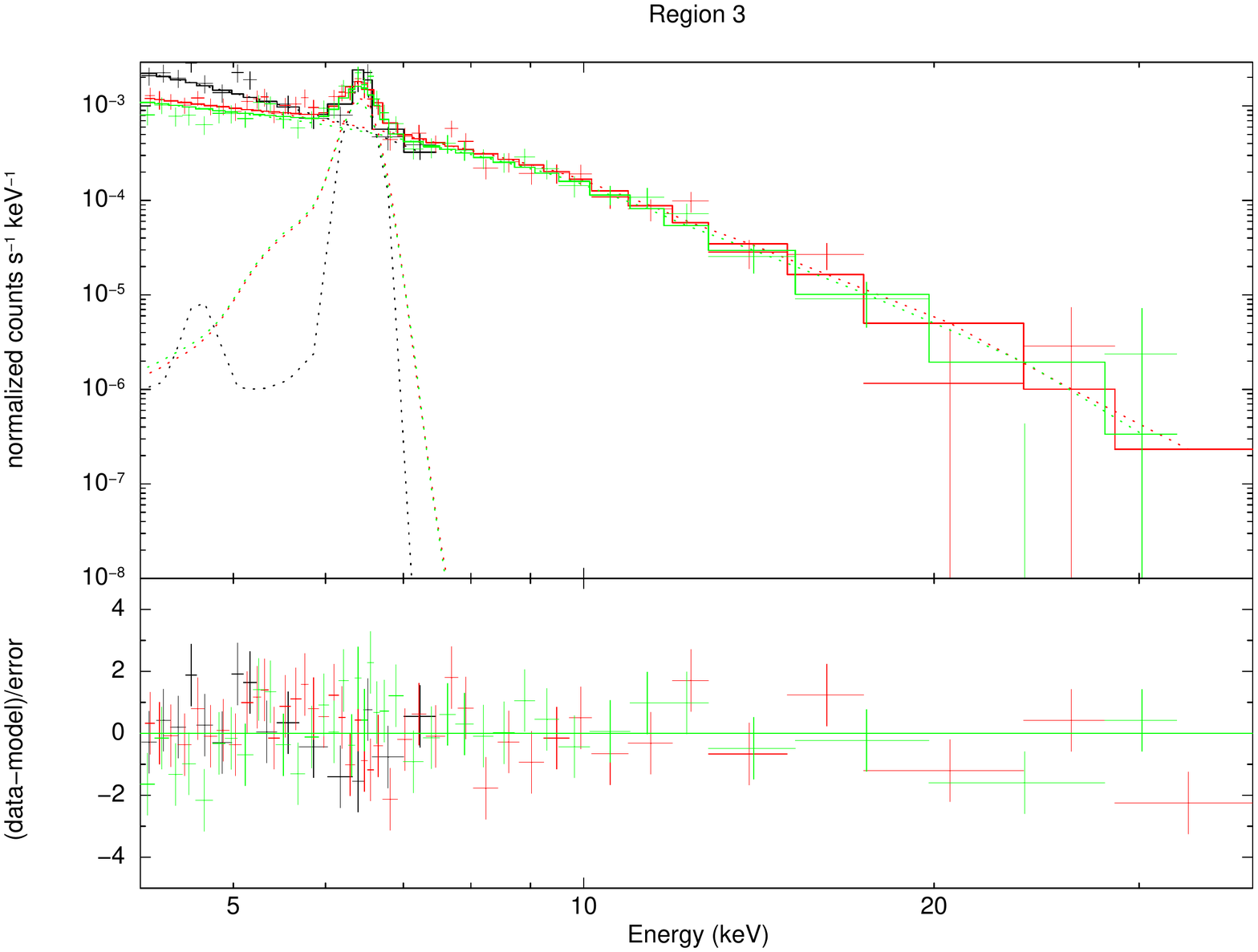}
    \includegraphics[width=0.3\columnwidth]{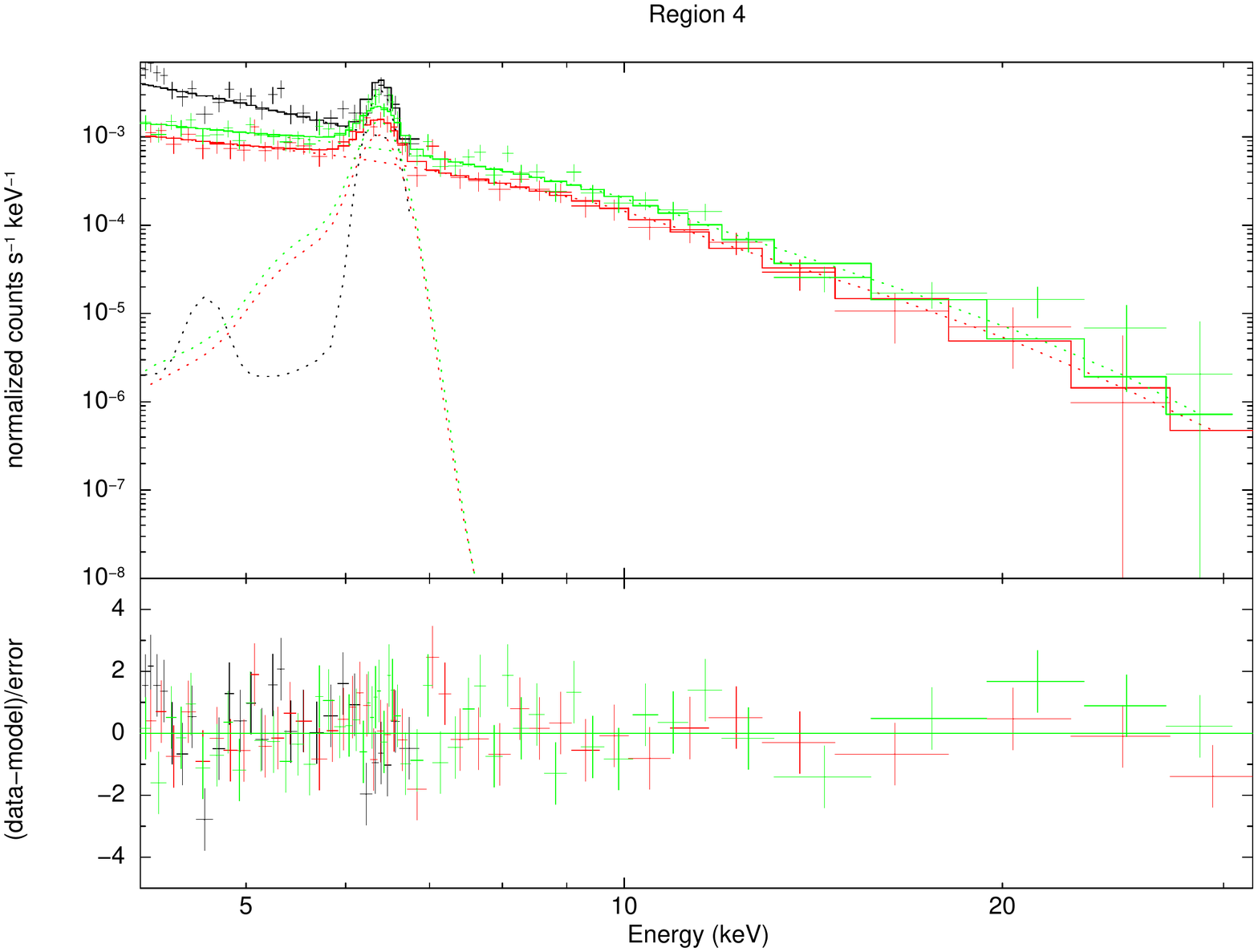}
    \includegraphics[width=0.3\columnwidth]{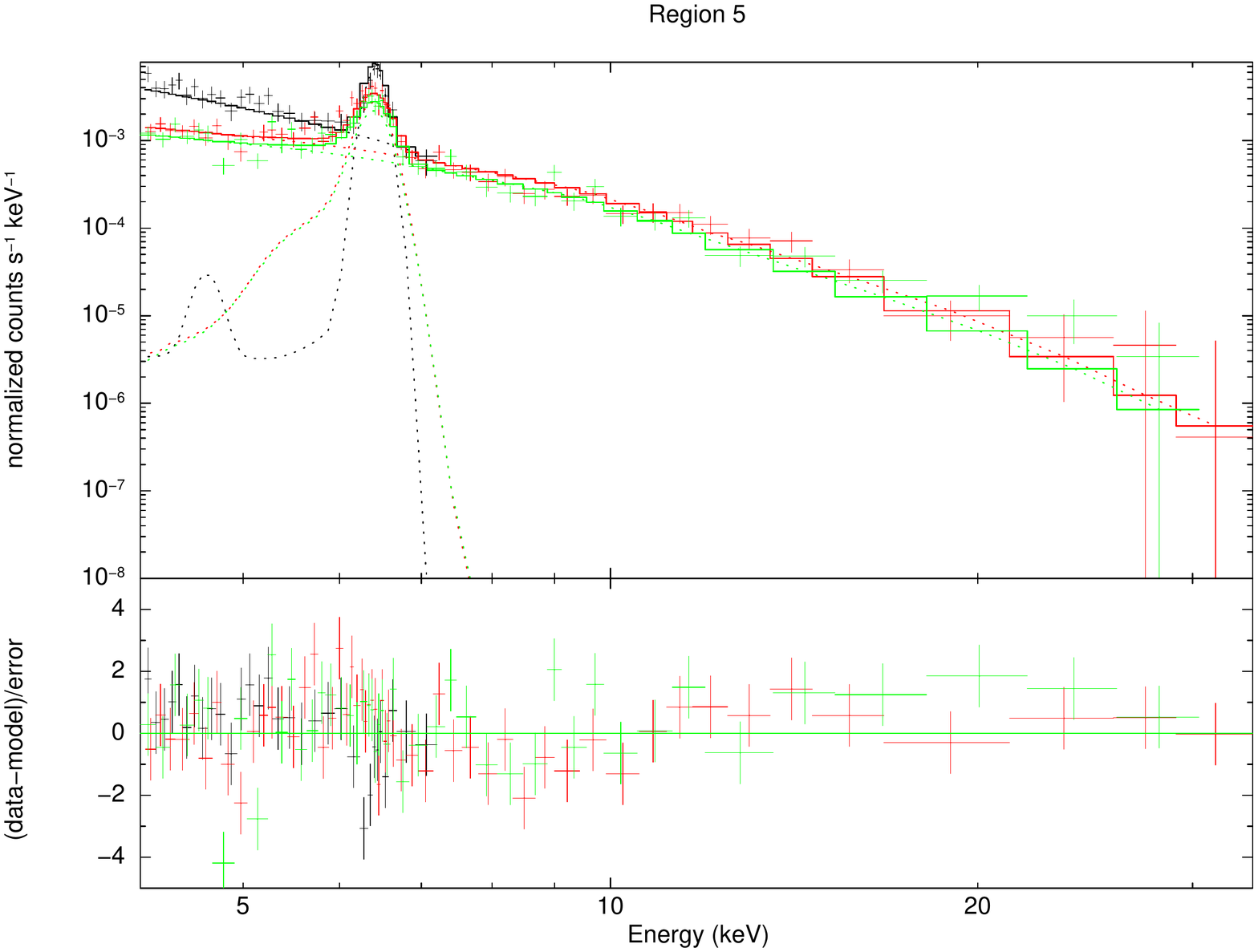}
    \includegraphics[width=0.3\columnwidth]{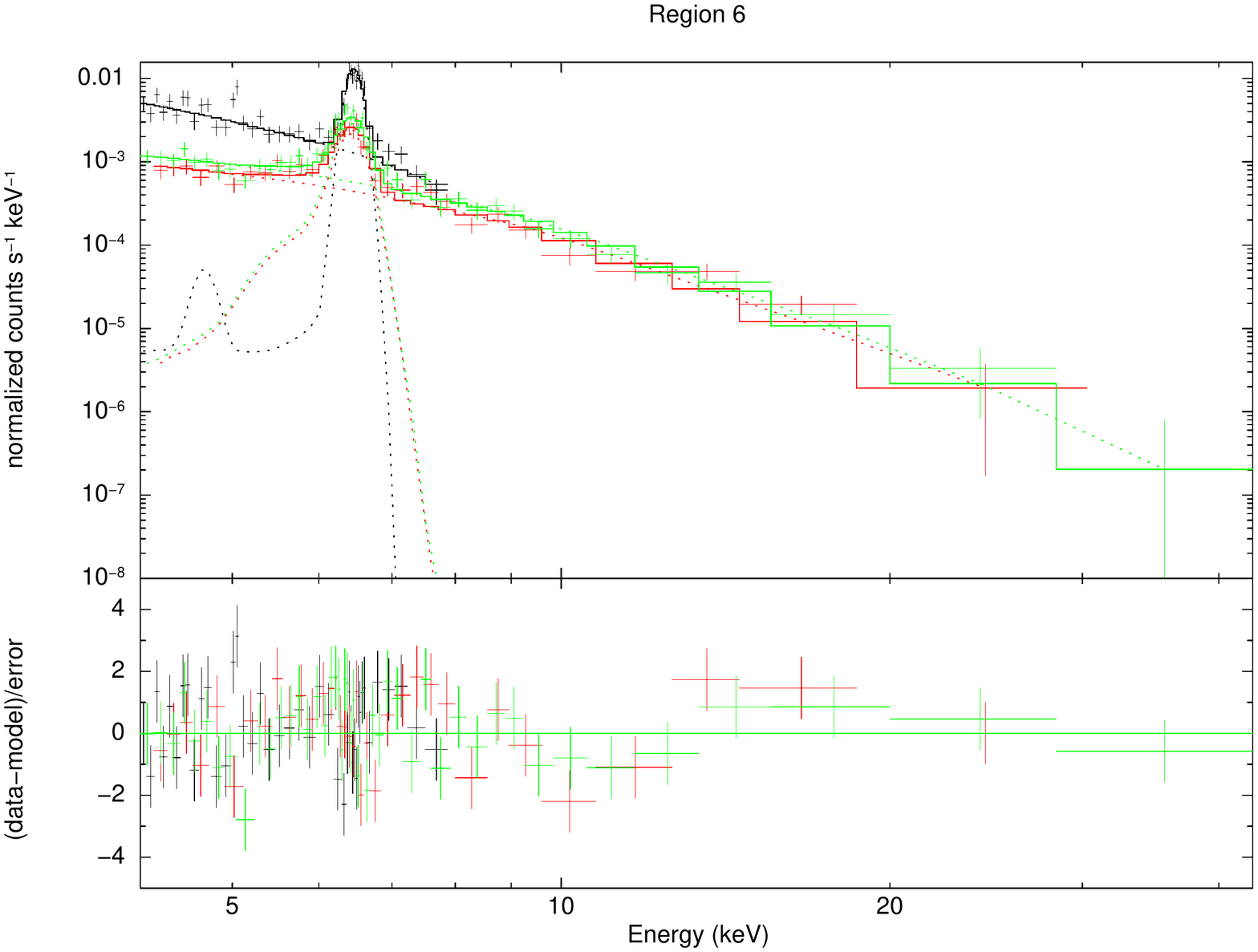}
    \includegraphics[width=0.3\columnwidth]{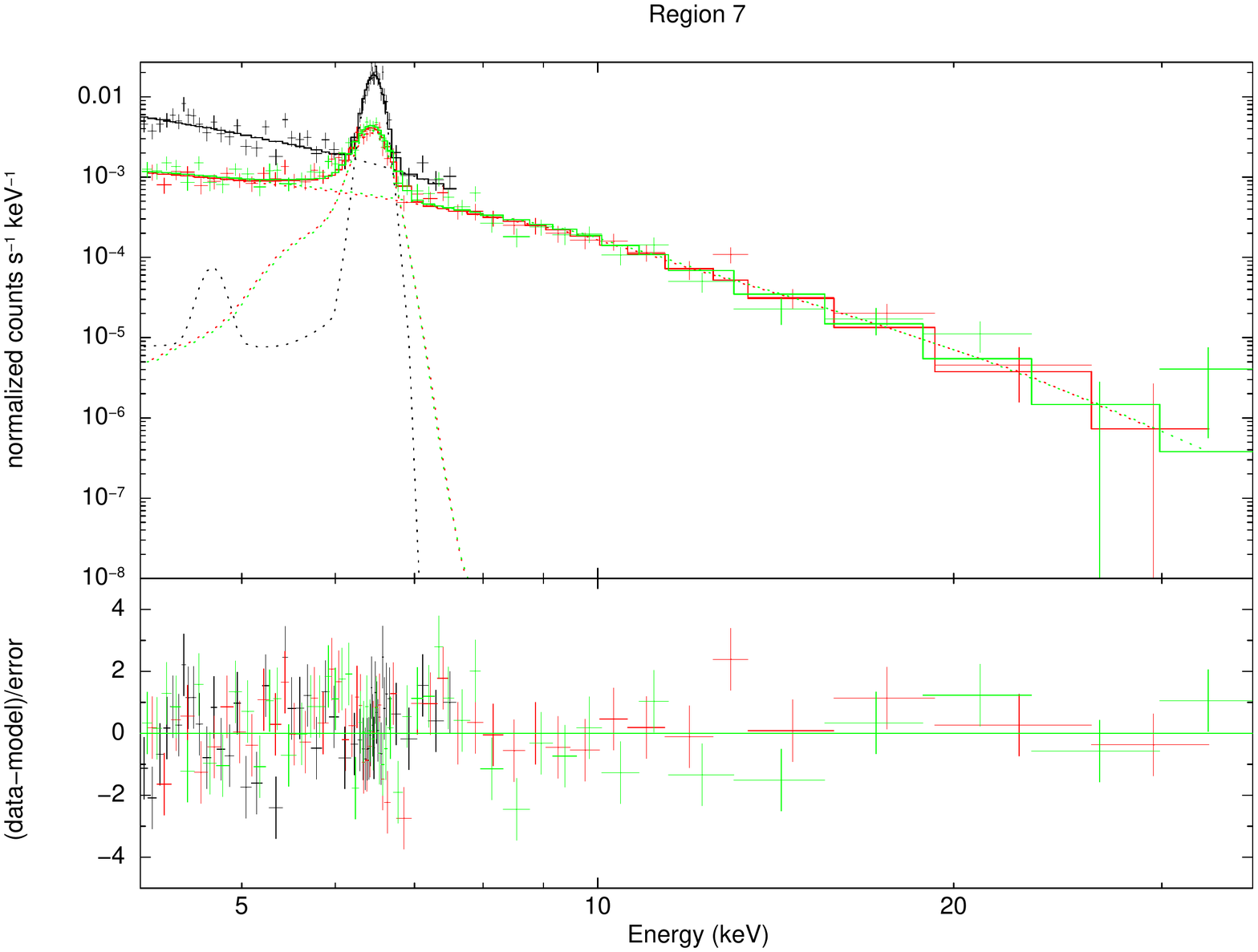}
    \includegraphics[width=0.3\columnwidth]{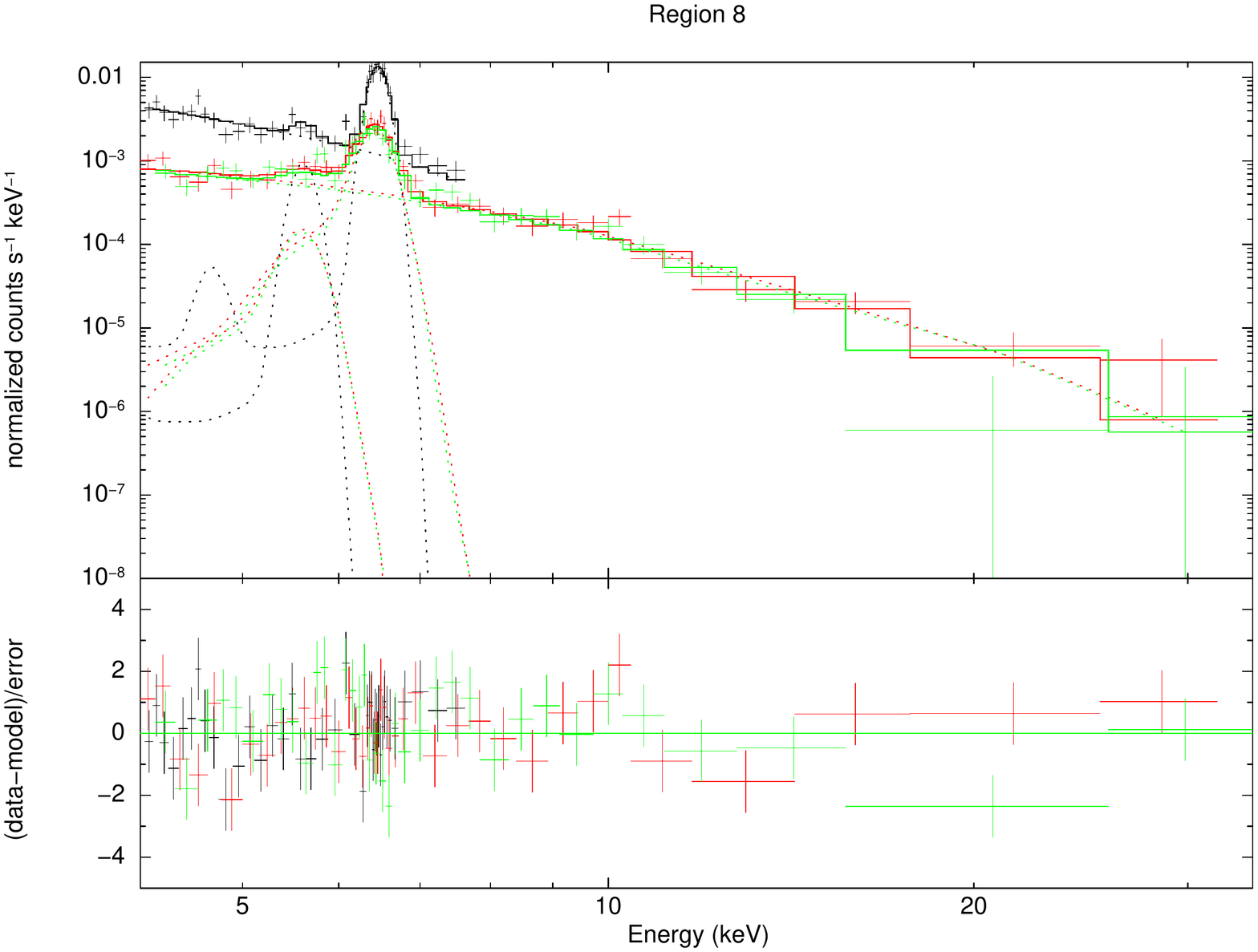}
    \includegraphics[width=0.3\columnwidth]{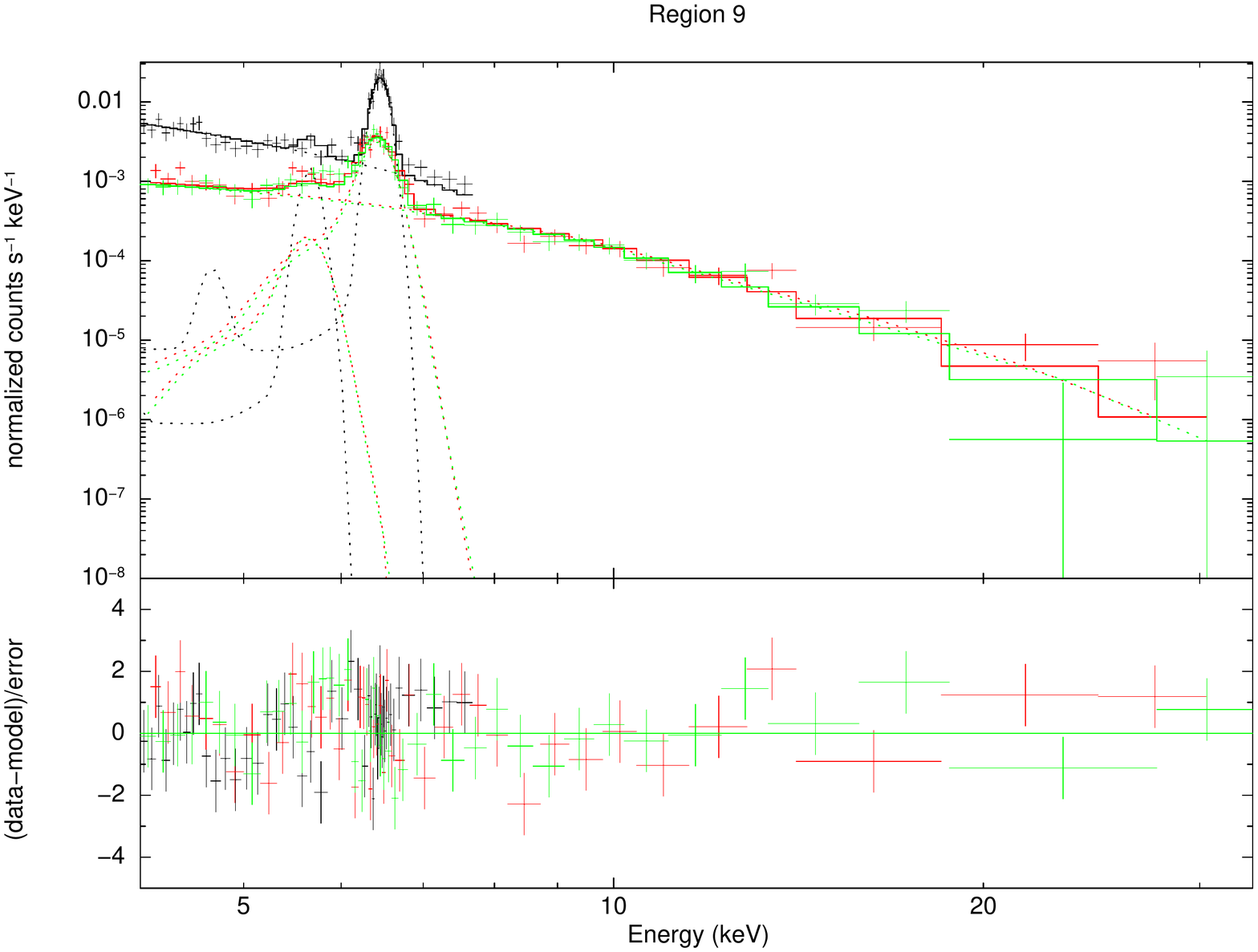}
    \includegraphics[width=0.3\columnwidth]{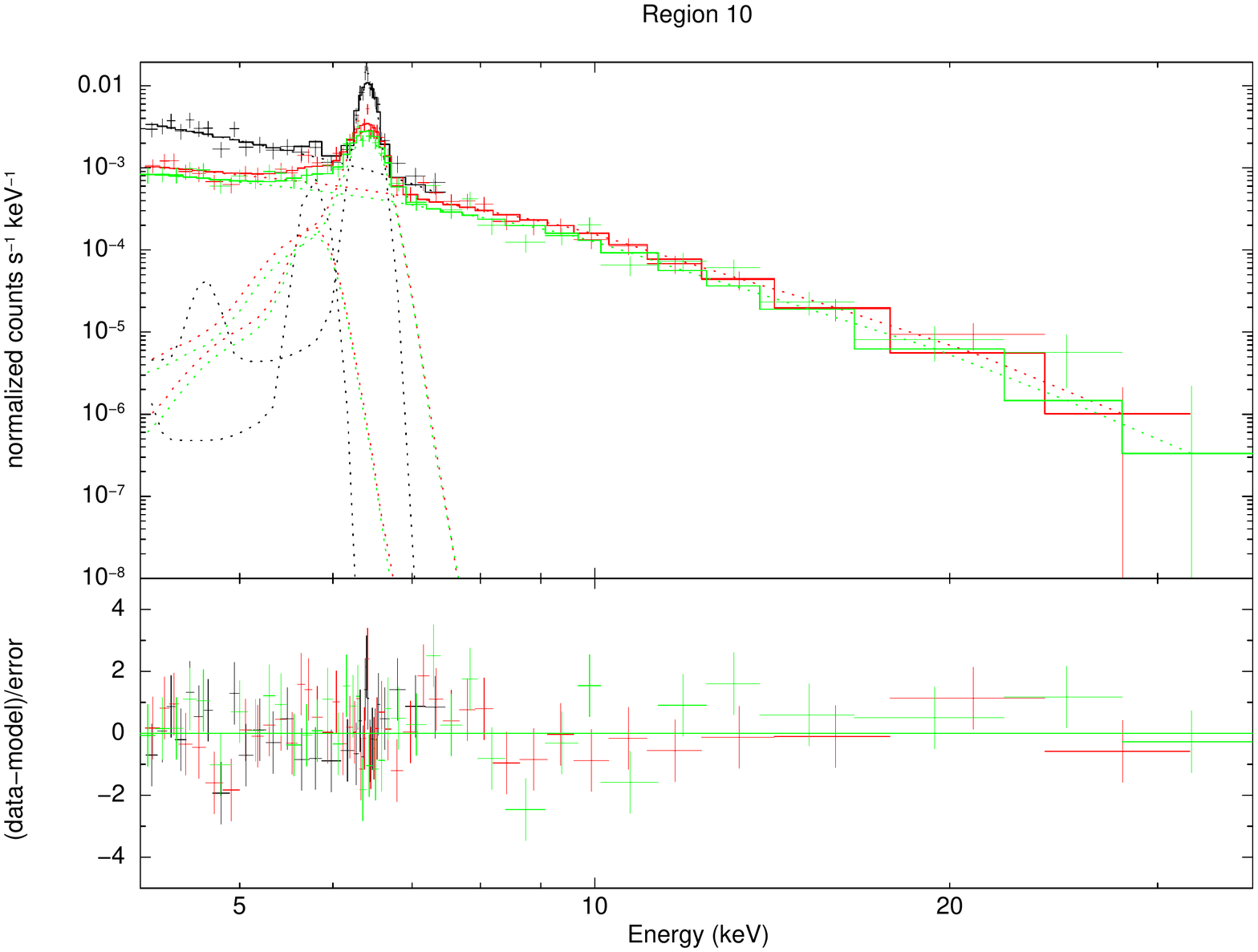}
    \includegraphics[width=0.3\columnwidth]{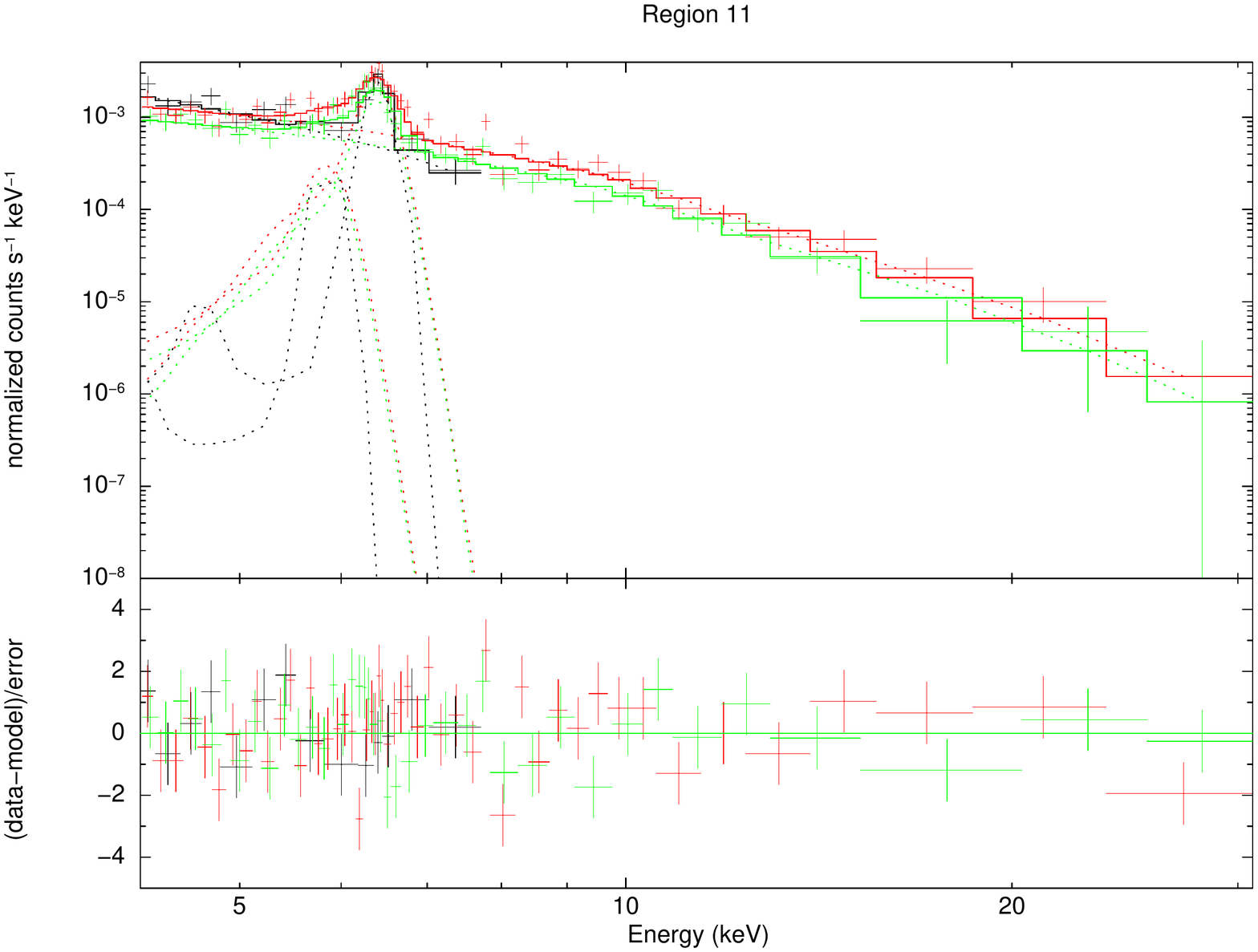}
    \caption{EPIC-pn (black), FPMA (red) and FPMB (green) spectra extracted from regions 1-11 (see Fig. \ref{fig:3to8vs8to20}) with the corresponding best-fit model and residual in the $4.1-30$ keV band (see Table \ref{tab:ecut} for the best-fit values).}
    \label{fig:1ziraonly}
\end{figure}

\begin{figure*}
    \centering
    \includegraphics[width=0.3\columnwidth]{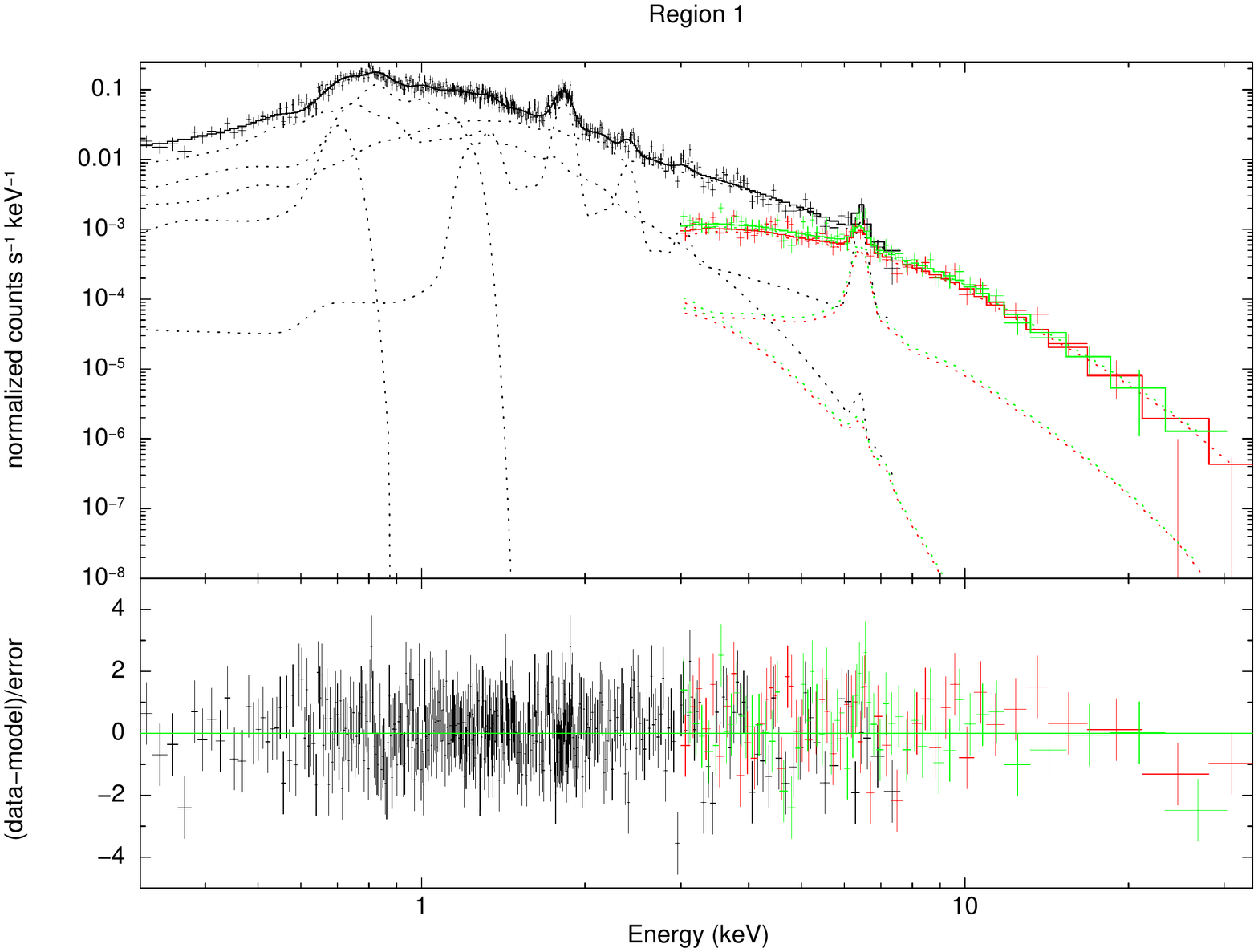}
    \includegraphics[width=0.3\columnwidth]{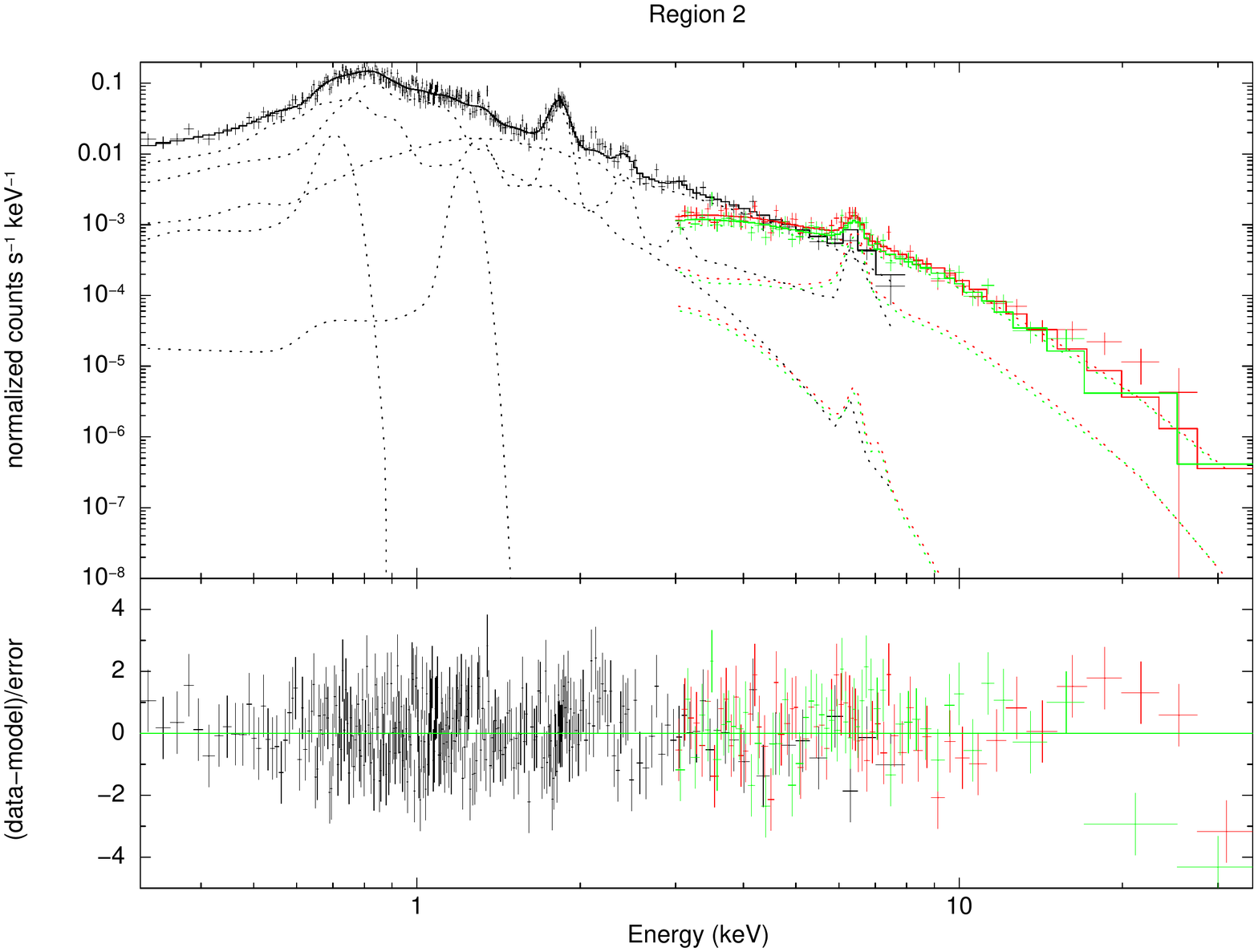}
    \includegraphics[width=0.3\columnwidth]{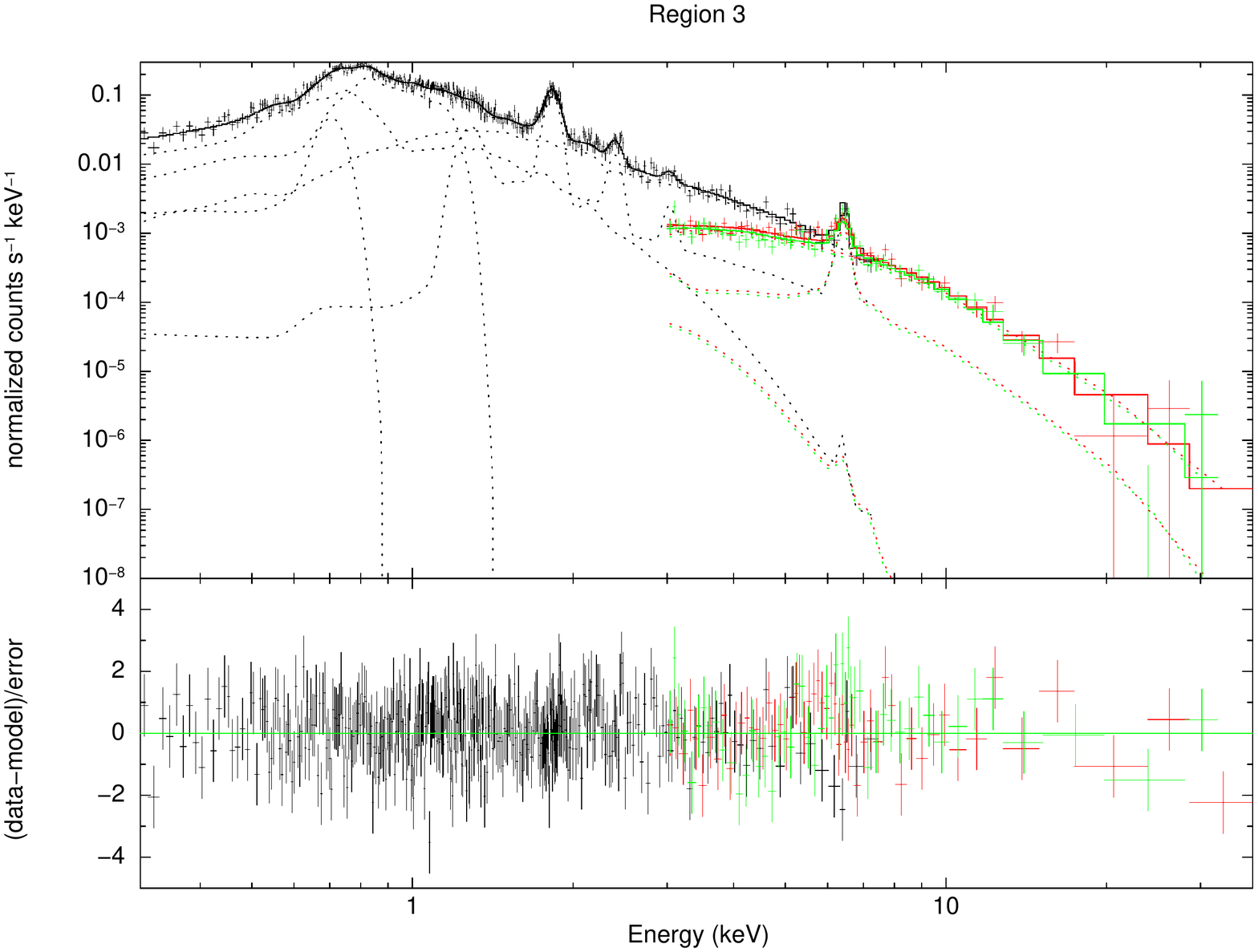}
    \includegraphics[width=0.3\columnwidth]{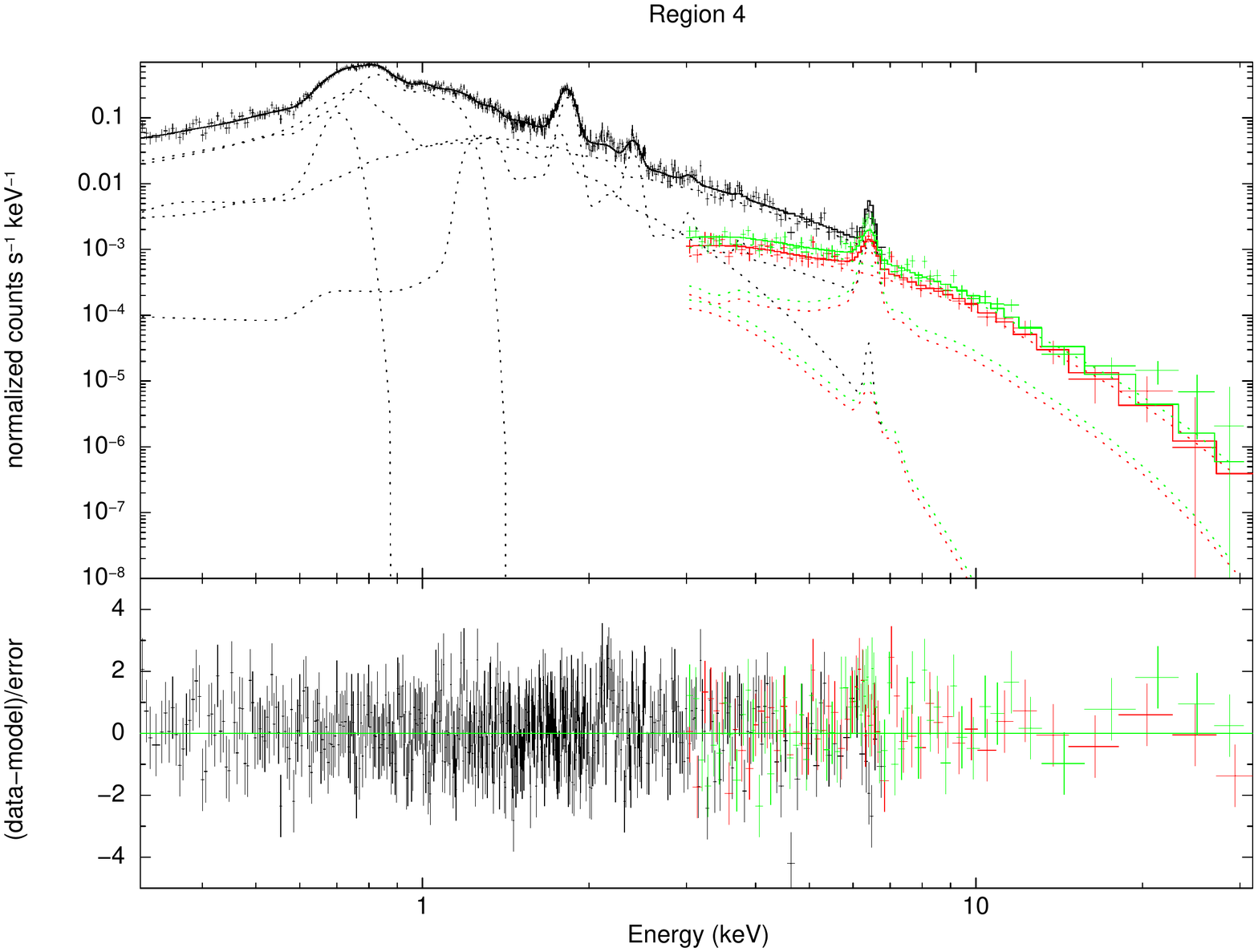}
    \includegraphics[width=0.3\columnwidth]{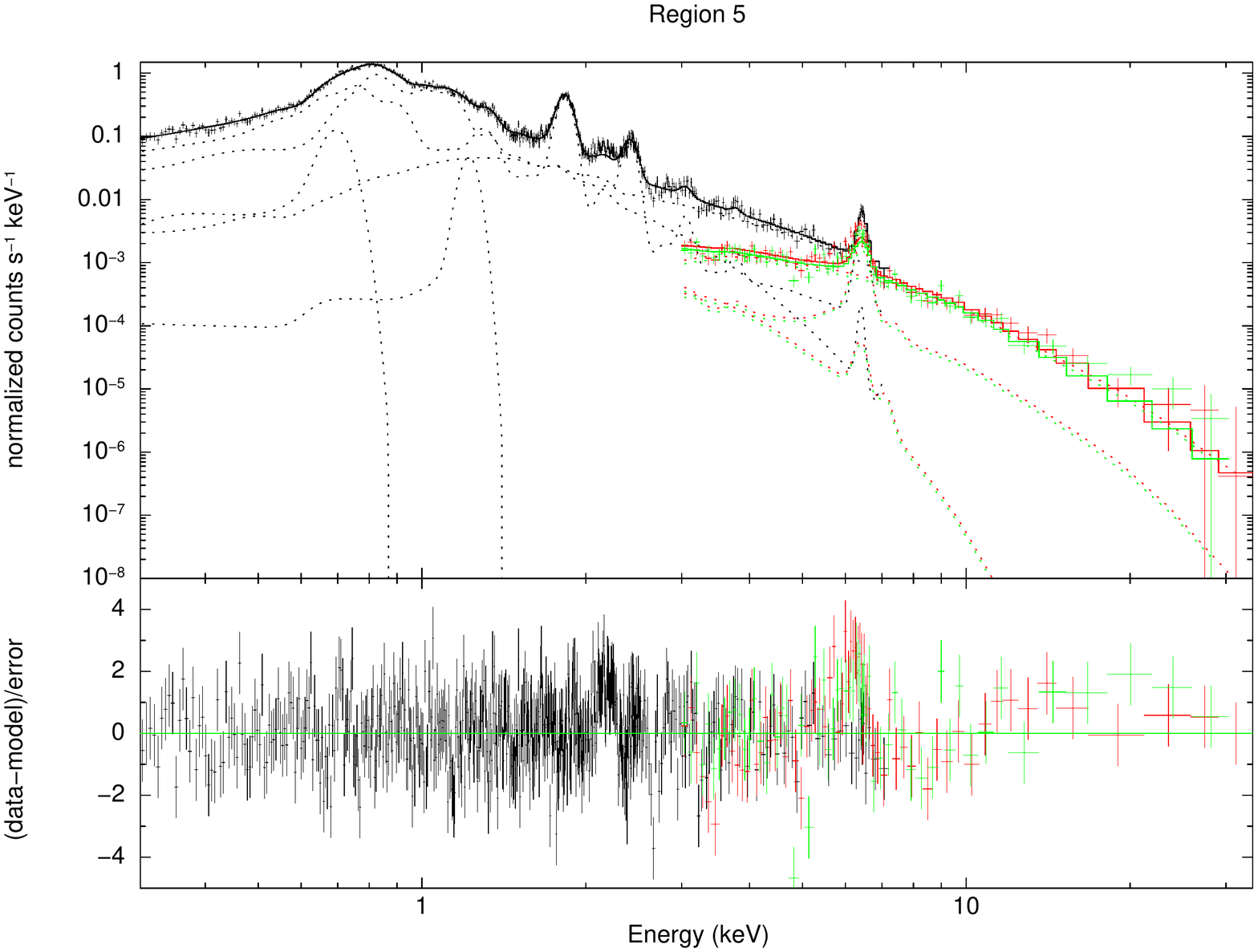}
    \includegraphics[width=0.3\columnwidth]{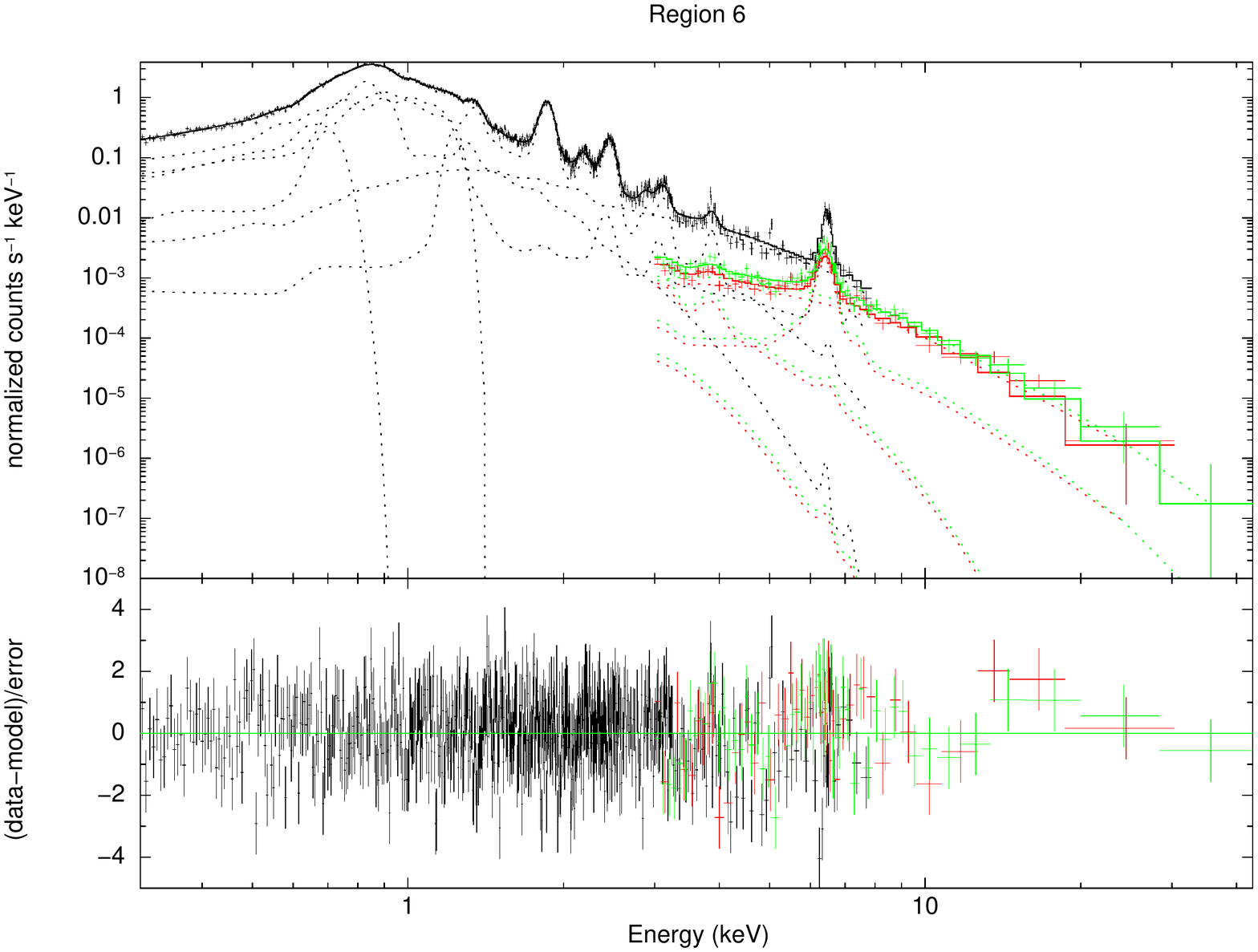}
    \includegraphics[width=0.3\columnwidth]{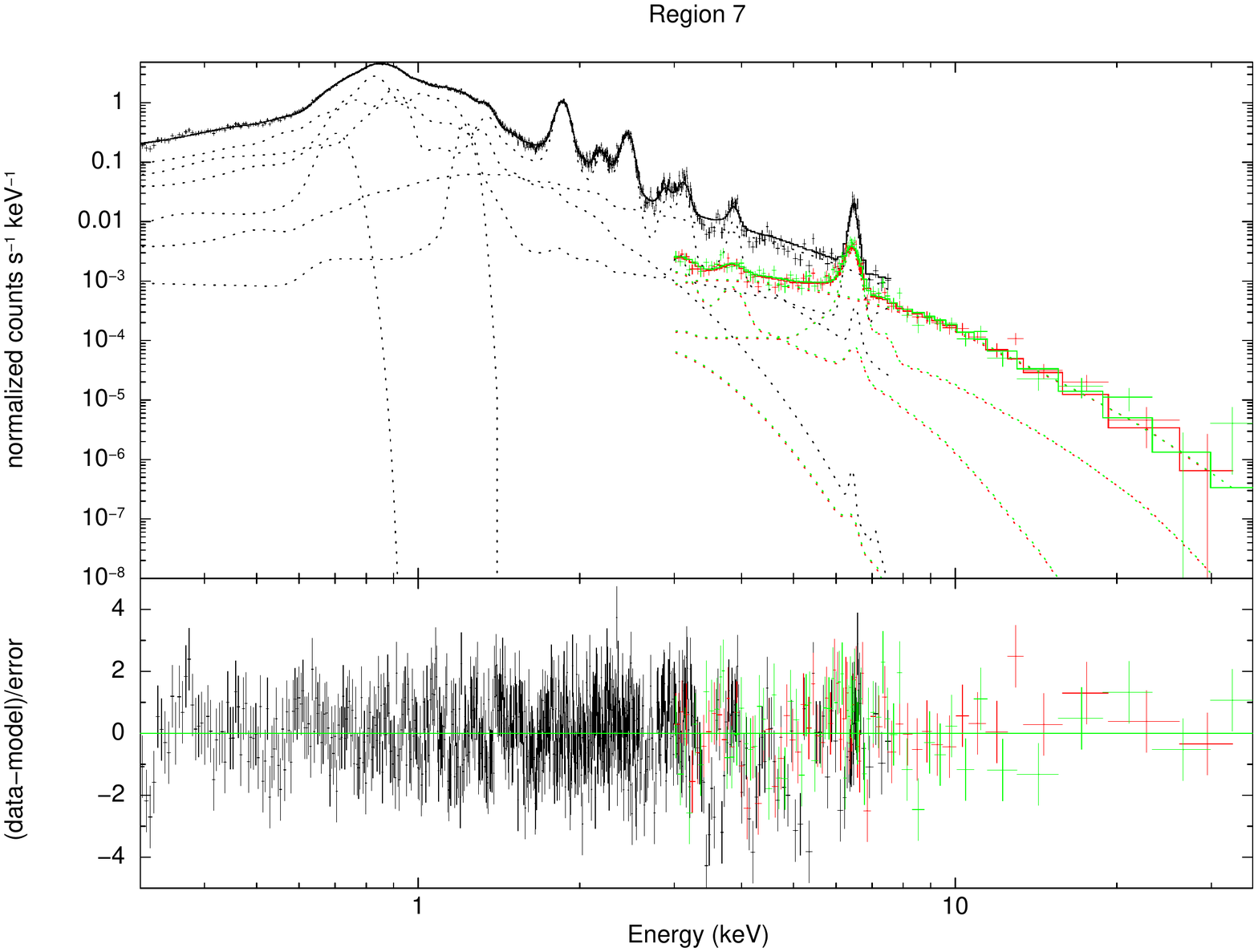}
    \includegraphics[width=0.3\columnwidth]{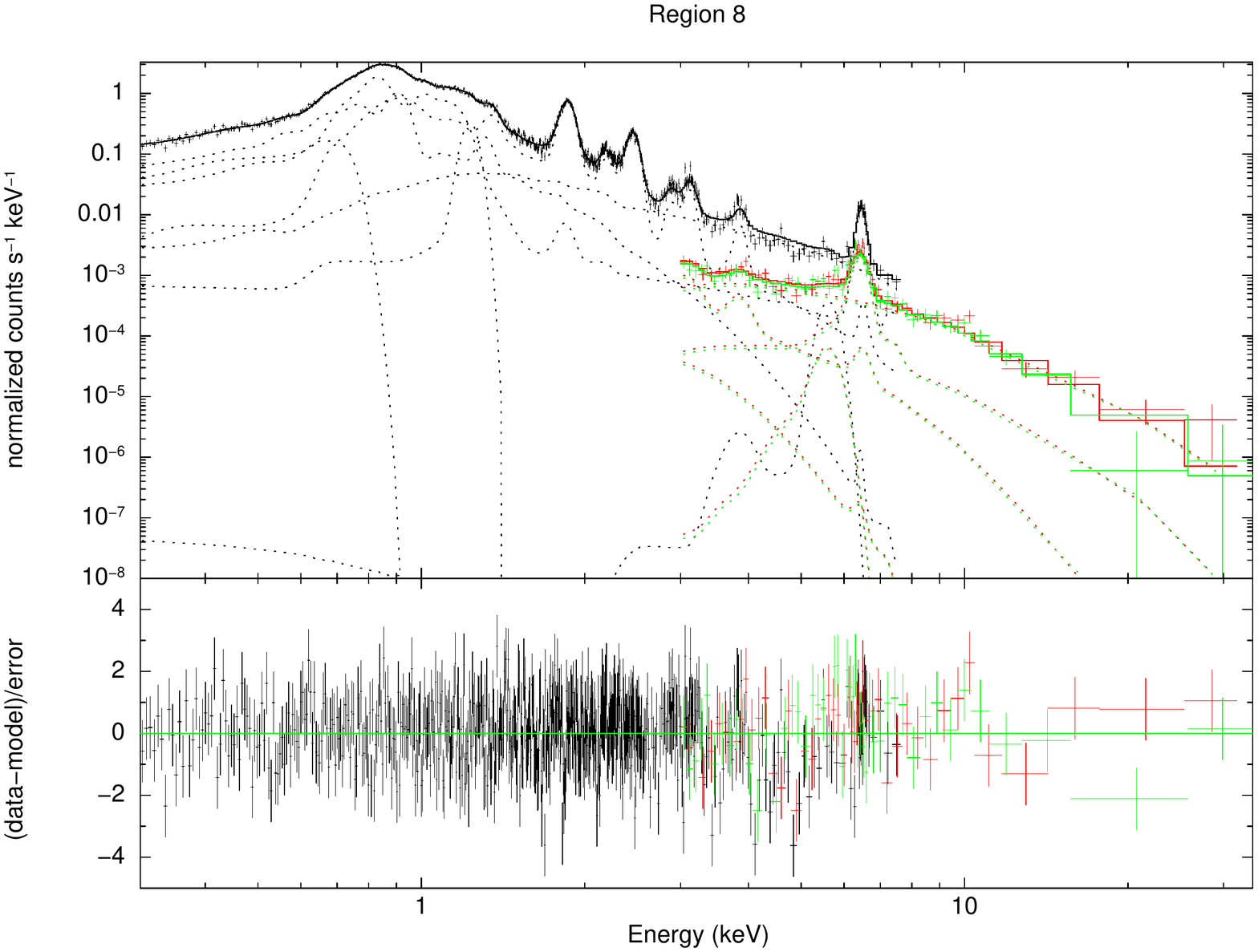}
    \includegraphics[width=0.3\columnwidth]{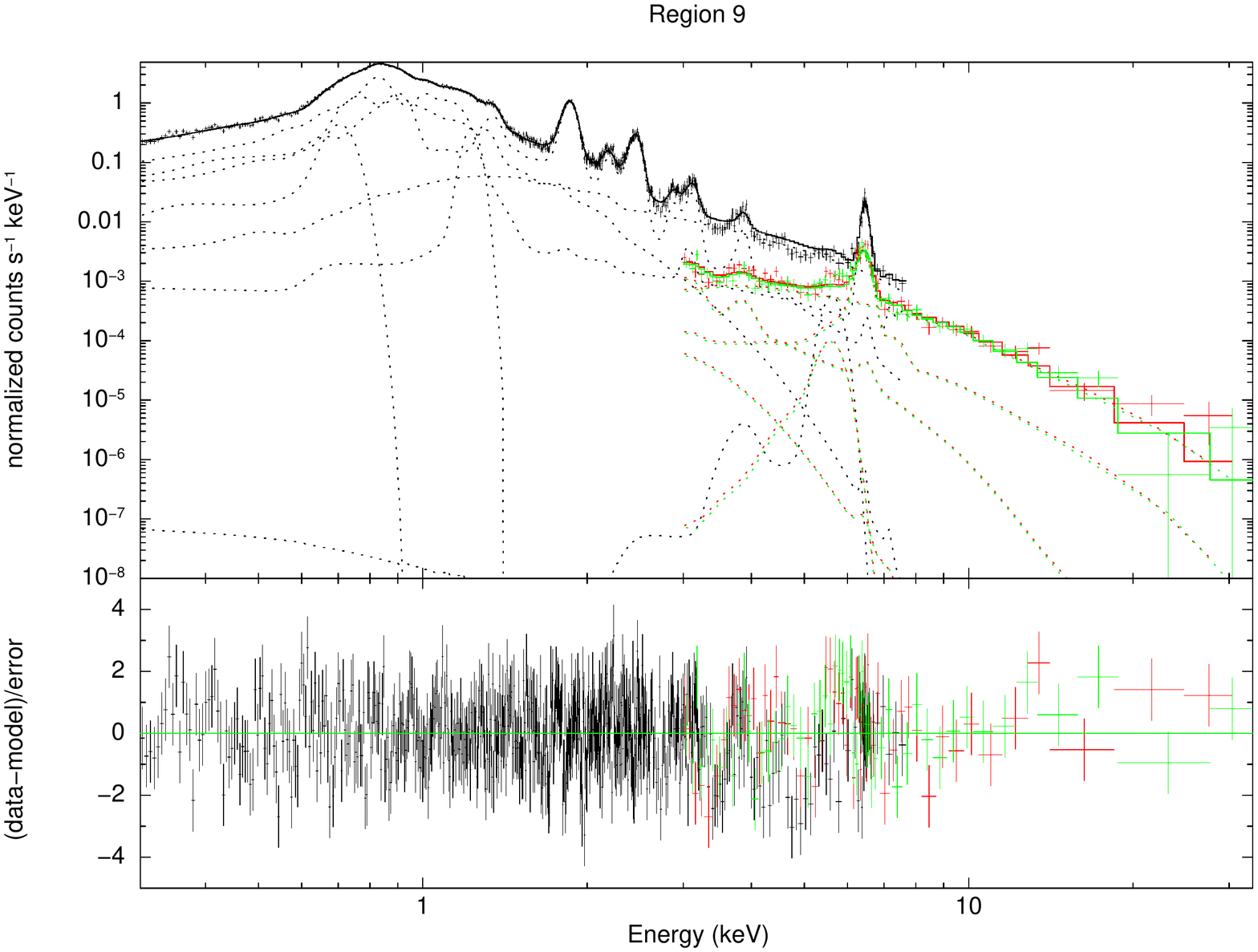}
    \includegraphics[width=0.3\columnwidth]{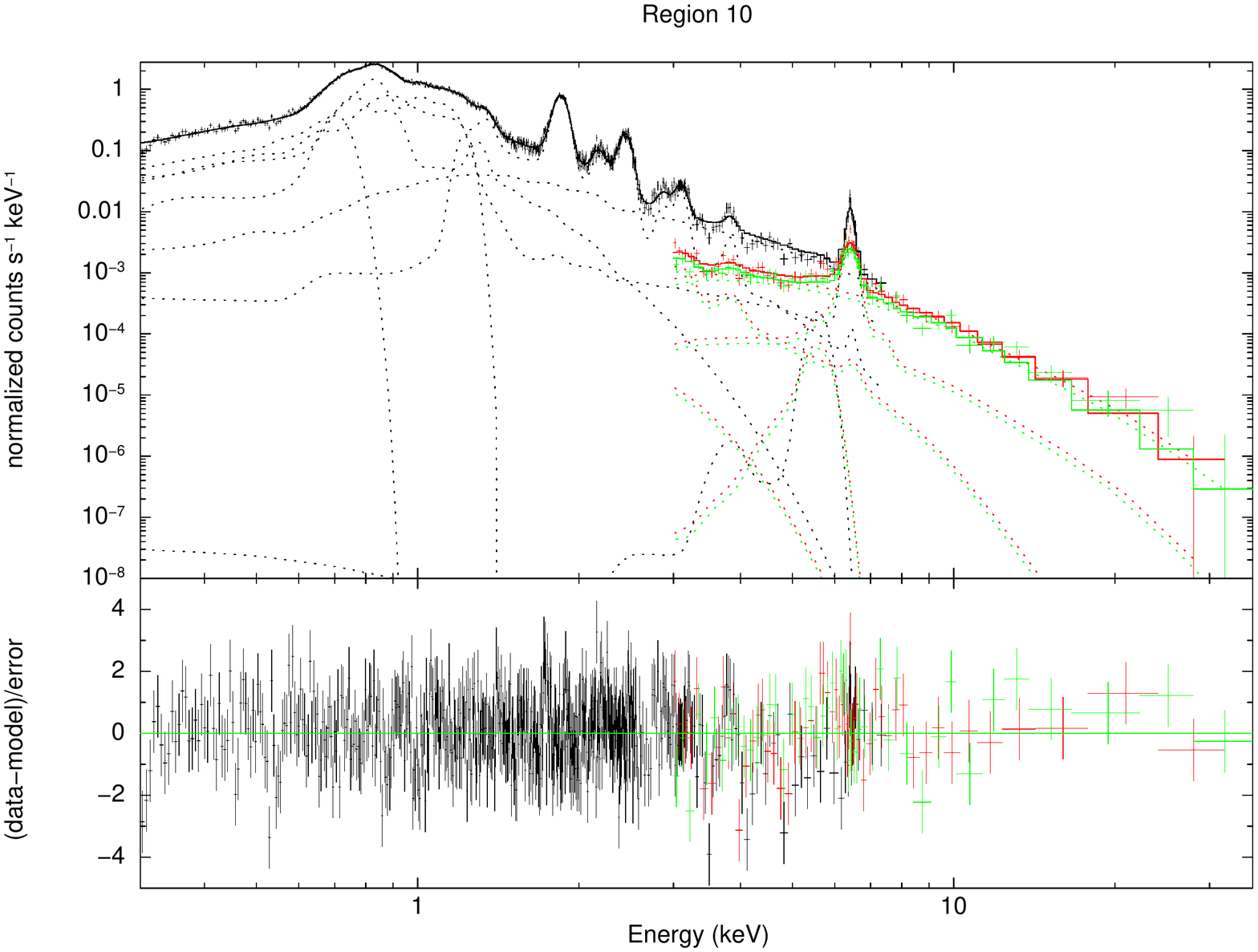}
    \includegraphics[width=0.3\columnwidth]{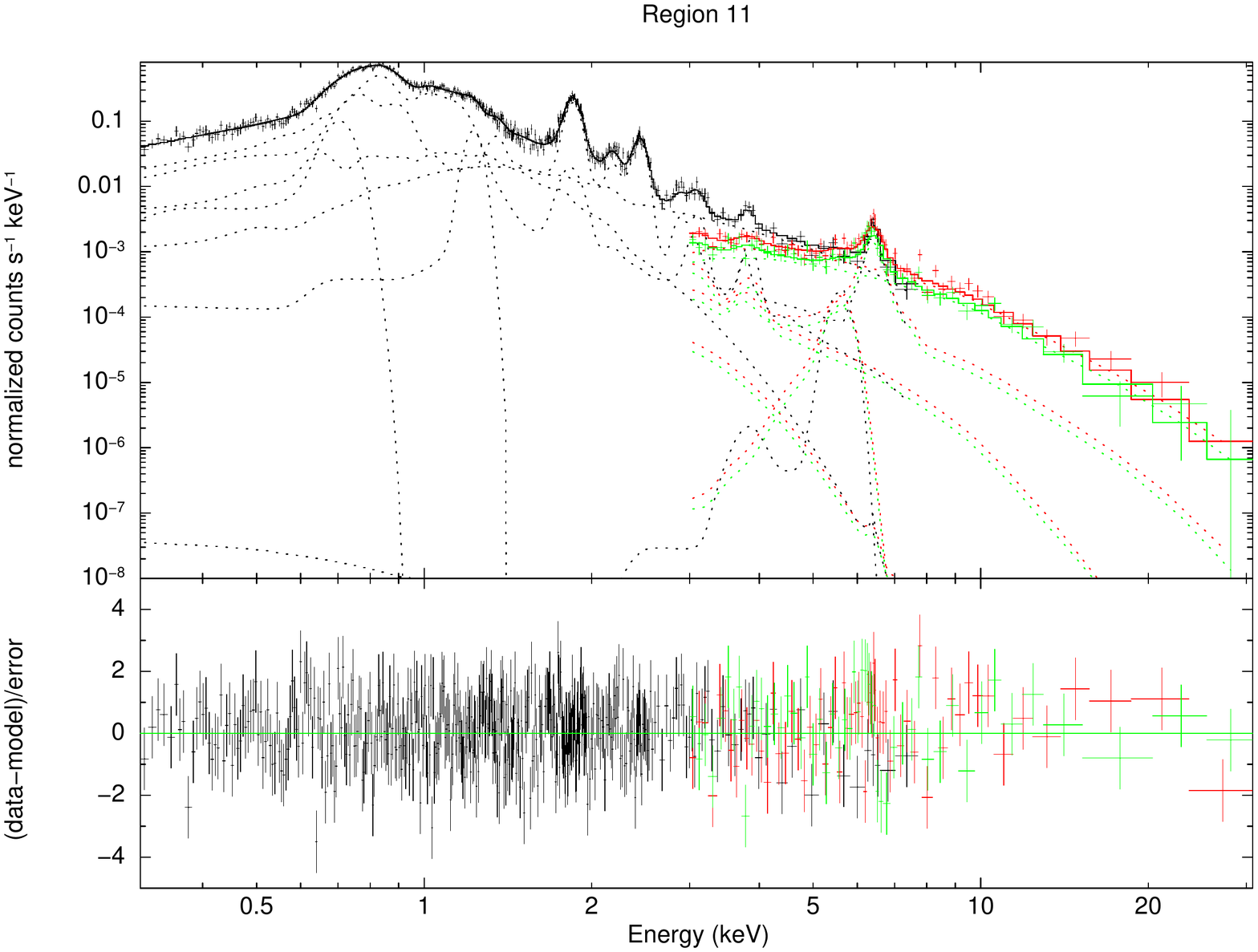}
    \caption{EPIC-pn (black), FPMA (red) and FPMB (green) spectra extracted from regions 1-11 (see Fig. \ref{fig:3to8vs8to20}) with the corresponding best-fit model and residual in the $0.3-30$ keV band.}
    \label{fig:8bvrneibf}
\end{figure*}

\clearpage

\section{Proper motion}
\label{app:pm}
To estimate the current velocity of the shock front in regions 1-11 (shown in Fig. \ref{fig:3to8vs8to20}), we considered the proper motion measurements by \citet{2022ApJ...926...84C} and \citet{Katsuda_2008}.
These measurements were obtained in small (with respect to our regions 1-11) regions at the shock front. For each region we considered the closest region(s) by \citet{2022ApJ...926...84C}, when available, and by \citet{Katsuda_2008} elsewhere, as shown in detail in Table \ref{tab:pm}. 
When more than one measurement of the proper motion were available, we considered their arithmetic mean. 
We did not find measurements of the proper motion in the areas of the shell corresponding to our region 1 and 10.

\begin{table}[!h]
    \centering
    \caption{Match between regions with proper motion measurements available in the literature, and regions from this work.}
    \label{tab:pm}
    \begin{tabular}{ccc}
    \hline\hline
        Region \# & \citet{2022ApJ...926...84C} Region \# & \citet{Katsuda_2008} Region \#\\
         \hline
        1& / &/\\
        2& 7 &/\\
        3& 8-9&/\\
        4& 9&/\\
        5&/&9-10-11\\
        6&/&13\\
        7&16&/\\
        8&1&/\\
        9&2&/\\
        10&/&/\\
        11&3-4&/\\
         \hline
    \end{tabular}

\end{table}


\bibliography{biblio}{}
\bibliographystyle{aasjournal}



\end{document}